\documentclass[preprint,twocolumn,epjc3]{svjourX}

\journalname{}

\input{packages-epjc.tex}

\newcommand{\tabt}[1]{\multicolumn{1}{c}{#1}}

\newcommand{\etal}{{\it et al.}}

\newcommand{\clicild}{\textsc{CLIC}\_\textsc{ILD}\xspace}

\newcommand{\marlin}{\textsc{Marlin}\xspace}
\newcommand{\geant}{\textsc{Geant4}\xspace}

\newcommand{\fastjet}{\textsc{FastJet}\xspace}
\newcommand{\higgsstrahlung}{\text{Higgsstrahlung}\xspace}

\newcommand{\gamgam}{\ensuremath{\PGg\PGg}\xspace}
\newcommand{\gghadrons}{\mbox{\ensuremath{\gamgam \to \text{hadrons}}}\xspace}

\newcommand{\roots }{\ensuremath{\sqrt{s}}\xspace}
\newcommand{\mZ }{\ensuremath{m_{\PZ}}\xspace}
\newcommand{\mW}{\ensuremath{m_{\PW}}\xspace}
\newcommand{\mH }{\ensuremath{m_{\PH}}\xspace}

\newcommand{\GeV}{\ensuremath{\text{GeV}}\xspace}

\newcommand{\TeV}{\ensuremath{\text{TeV}}\xspace}

\newcommand{\fbinv}{\ensuremath{\text{fb}^{-1}}\xspace}

\newcommand{\gHZZ}{\ensuremath{g_{\PH\PZ\PZ}}\xspace}
\newcommand{\gHXX}{\ensuremath{g_{\PH X X}}\xspace}
\newcommand{\gHWW}{\ensuremath{g_{\PH\PW\PW}}\xspace}

\newcommand{\BR}{\ensuremath{\text{BR}}\xspace}
\newcommand{\rootsprime }{\ensuremath{\sqrt{s^\prime}}\xspace}
\newcommand{\pT}{\ensuremath{p_\mathrm{T}}\xspace}

\newcommand{\mrec}{\ensuremath{m_\mathrm{rec}}\xspace}

\newcommand{\mqq}{\ensuremath{m_{\PQq\PAQq}}\xspace}
\newcommand{\mqqp}{\ensuremath{m^\prime_{\PQq\PAQq}}\xspace}
\newcommand{\cosQ}{\ensuremath{|\cos\theta_{\PQq}|}}

\newcommand{\thetamis}{\ensuremath{\theta_{\text{mis}}}}

\newcommand{\epem}{\ensuremath{\Pep\Pem}\xspace}
\newcommand{\mpmm}{\ensuremath{\PGmp\PGmm}\xspace}  
\newcommand{\tptm}{\ensuremath{\PGtp\PGtm}\xspace} 
\newcommand{\qq}{\ensuremath{\PQq \PAQq}\xspace}

\newcommand{\qqbar}{\ensuremath{\PQq \PAQq}\xspace} 
\def \qq {\qqbar}
 
\newcommand{\lplm}{\ensuremath{\Plp\Plm}\xspace}

\renewcommand{\Pl}{\ell}

\begin{document}

\title{\boldmath\bf Model-Independent Measurement of the e$^{\text{+}}$e$^\text{--}$ $\to$ HZ  Cross Section at a Future e$^{\text{+}}$e$^\text{--}$ Linear Collider using Hadronic Z Decays}

\author{M. A. Thomson\thanksref{Cam}}

\institute{Cavendish Laboratory, University of Cambridge, JJ Thomson Avenue, Cambridge, CB3\,0HE, UK \label{Cam}}

\date{9th September 2015}

\maketitle

\begin{abstract}
  A future $\epem$ collider, such as the ILC or CLIC, would allow the 
  Higgs sector to be probed with a precision significantly beyond that achievable at the
  High-Luminosity LHC. A central part of the Higgs programme at an $\epem$ collider is the model-independent determination of the absolute Higgs couplings to
  fermions and to gauge bosons. Here the measurement of the $\epem\to\PH\PZ$ \higgsstrahlung cross section, using 
  the recoil mass technique, sets the absolute scale for all Higgs coupling measurements. Previous studies have considered
  $\sigma(\epem\to\PH\PZ)$ with  $\PZ\to\lplm$, where $\Pl = \Pe,\,\PGm$.
  In this paper it is shown for the first time that a near model-independent recoil mass technique 
  can be extended to the hadronic decays of the $\PZ$ boson. Because the branching ratio for 
  $\PZ\to\qq$ is approximately ten times greater than for $\PZ\to\lplm$, this method is statistically more powerful than 
  using the leptonic decays. For an integrated luminosity of 500\,\fbinv at a centre-of-mass energy of $\roots=350\,\GeV$ at CLIC,
  $\sigma(\epem\to\PH\PZ)$ can be measured to $\pm1.8\,\%$ using the hadronic recoil mass technique. A similar precision is found for
  the ILC operating at $\roots=350\,\GeV$. The centre-of-mass dependence of this measurement technique is discussed, arguing for the
  initial operation of a future linear collider at just above the top-pair production threshold.     
\end{abstract}

\section{Introduction}
A future $\epem$ collider, such as the Compact Linear Collider (CLIC)~\cite{bib:CLIC_CDR} or the International Linear Collider (ILC)~\cite{bib:ILC_TDR}, 
would be complementary to the Large Hadron Collider (LHC) and the High-Luminosity LHC (HL-LHC), providing tests of beyond the 
Standard Model (SM) physics scenarios through a broad programme of highly precise measurements. A central part of the linear collider physics 
programme is the precise study of the properties of the Higgs boson. The LHC and HL-LHC will provide an impressive range of Higgs physics measurements, 
establishing the general properties of the Higgs boson, such as its mass and spin. The LHC will also provide measurements of the {\it product} of the Higgs 
production rate and Higgs decay branching fractions into different final states. Current estimates suggest that {\it ratios} of couplings can be 
measured to 2\,\% -- 7\,\% (depending on the final state) with 3000\,fb$^{-1}$ of data~\cite{bib:SnowmassHiggs}. A number of recent studies,
see for example~\cite{bib:SnowmassHiggs,bib:Wells}, have indicated that modifications of the Higgs couplings due to beyond the SM (BSM) physics 
are almost always less than 10\,\% and can be as small as 1\,\% -- 2\,\% in a number of models. 
 
An $\epem$ collider would be a unique facility for precision Higgs physics~\cite{bib:ILC_Physics_TDR, bib:CLIC_Snowmass}, 
providing measurements of the Higgs boson branching ratios that may be an order of magnitude more precise than those achievable at the HL-LHC. Such measurements
may be necessary to reveal BSM effects in the Higgs sector. Moreover, an $\epem$ collider provides the opportunity to make a number of unique measurements including: 
i) absolute measurements of Higgs couplings, rather than ratios; ii) a precise measurement of possible decays to invisible 
(long-lived neutral) final states; and iii) a $<10\,\%$ measurement of the total Higgs decay width, $\Gamma_{\PH}$. In addition, an $\epem$ collider 
operating at 1\,$\TeV$ or above, for example CLIC or an upgraded ILC, would have sensitivity  to the Yukawa coupling of the top quark to the Higgs boson
and the Higgs self-coupling parameter $\lambda$, thus providing a direct probe of the Higgs potential.

This paper presents the first detailed study of the potential of making a model-independent measurement of $\gHZZ$ from the recoil mass distribution in
$\epem\to\PH\PZ$ with $\PZ\to\qqbar$, denoted as $\PH\PZ(\PZ\to\qqbar)$. The studies were initially performed in the context of the CLIC accelerator operating at $\roots = 350\,\GeV$.  The studies were repeated for the ILC operating at the same energy and for CLIC at $\roots=250\,\GeV$ and $\roots=420\,\GeV$. 

\subsection{Higgs production in $e^+e^-$ collisions}

In $\epem$ collisions at $\roots$ = 250 -- 500\,GeV, the two main Higgs production mechanisms are 
the \higgsstrahlung process, $\Pep\Pem\to\PH\PZ$, and the $\PW\PW$-fusion process, $\Pep\Pem\to\PH\PGne\PAGne$, shown in
Fig.~\ref{fig:eezh}. For $\mH \sim 125\,$GeV, the cross section for the $s$-channel \higgsstrahlung process is maximal close to $\roots = 250\,\GeV$, whereas the
cross section for the $t$-channel $\PW\PW$-fusion process increases with centre-of-mass energy, as indicated in Tab.~\ref{tab:higgsprod}.

\begin{figure}[htb]
\unitlength = 1mm
\vspace{6mm}
\centering
\begin{fmffile}{eezh}
\begin{fmfgraph*}(25,20)
\fmfstraight
\fmfleft{i1,i2}
\fmfright{o1,o2}
\fmflabel{$\Pem$}{i1}
\fmflabel{$\Pep$}{i2}
\fmflabel{$\PZ$}{o2}
\fmflabel{$\PH$}{o1}
\fmf{photon,tension=1.0,label=$\PZ$}{v1,v2}
\fmf{fermion,tension=1.0}{i1,v1,i2}
\fmf{photon,tension=1.0}{o2,v2}
\fmf{dashes,tension=1.0}{v2,o1}
\fmfdot{v1}
\fmfdot{v2}
\end{fmfgraph*}
\end{fmffile}
\hspace{10mm}
\begin{fmffile}{eevvh}
\begin{fmfgraph*}(25,20)
\fmfstraight
\fmfleft{i1,i2}
\fmfright{o1,oh,o2}
\fmflabel{$\Pem$}{i1}
\fmflabel{$\Pep$}{i2}
\fmflabel{$\PAGne$}{o2}
\fmflabel{$\PH$}{oh}
\fmflabel{$\PGne$}{o1}
\fmf{fermion, tension=2.0}{i1,v1}
\fmf{fermion, tension=1.0}{v1,o1}
\fmf{fermion, tension=1.0}{o2,v2}
\fmf{fermion, tension=2.0}{v2,i2}
\fmf{photon, lab.side=right,lab.dist=1.5,label=$\PW$,tension=1.0}{v1,vh}
\fmf{photon, lab.side=right, lab.dist=1.5,label=$\PW$,tension=1.0}{vh,v2}
\fmf{dashes, tension=1.0}{vh,oh}
\fmfdot{vh}
\end{fmfgraph*}
\end{fmffile}

\vspace{3mm}
\caption{
Feynman diagrams for: (left) the \higgsstrahlung process $\Pep\Pem\to\PH\PZ$, which dominates at $\roots$ = 250\,GeV; and (right) the $\PW\PW$-fusion process $\epem\to\PGne\PAGne\PH$, which dominates at $\roots$ > 500\,GeV. \label{fig:eezh}
}
\end{figure}

The total $\PH\PZ$ cross section is proportional to the square of the coupling between the Higgs and $\PZ$ bosons, $\gHZZ$,
\begin{equation*}
     \sigma(\epem\to\PH\PZ)  \propto \gHZZ^2 \,,
\end{equation*}
and the cross sections for the exclusive final-state decays $\PH\to X\bar{X}$ can be expressed as
\begin{align*}
     \sigma(\epem\to\PH\PZ) \times \BR(\PH\to X\bar{X}) &\propto \frac{\gHZZ^2 \times \gHXX^2}{\Gamma_{\PH}} \\ 
     \sigma(\epem\to\PH\PGne\PAGne) \times \BR(\PH\to X\bar{X}) &\propto \frac{\gHWW^2 \times \gHXX^2}{\Gamma_{\PH}}\,.
\end{align*}
Once $\gHZZ$ has been determined in a  model-independent manner, the ratio of the \higgsstrahlung and $\PW\PW$-fusion cross sections for the same exclusive Higgs boson final state yields $\gHWW$. Subsequently, the measurement of $\sigma(\epem\to\PH\PGne\PAGne)$ $\times$ $\BR(\PH\to\PW\PW^*)$, which depends on $\gHWW^4/\Gamma_{\PH}$, provides a 
determination of $\Gamma_{\PH}$. At this point all measurements of exclusive Higgs decays provide absolute and model-independent  determinations of the 
relevant coupling(s). The determination of $\gHZZ$ from the recoil mass distribution in $\Pep\Pem\to\PH\PZ$ lies at the heart of this scientific programme.

\begin{table*}[htb]\centering
 \begin{tabular}{lccrrr}
   \toprule 
                                  & $\epem$ polarisation     & \tabt{$\roots$ =} & \tabt{250\,GeV} & \tabt{350\,GeV} & \tabt{500\,GeV} \\ \midrule
    $\sigma(\Pep\Pem\to\PH\PZ)$    &  unpolarised  & &     211\,fb         & 134\,fb           & 65\,fb         \\
    $\sigma(\Pep\Pem\to\PH\PGne\PAGne)$  &   unpolarised & & 21\,fb          & 34\,fb         & 72\,fb       \\  \midrule
    $\sigma(\Pep\Pem\to\PH\PZ)$    &  (-0.8,\,+0.3)  & &       318\,fb         & 198\,fb           & 96\,fb         \\
    $\sigma(\Pep\Pem\to\PH\PGne\PAGne)$  &   (-0.8,\,+0.3) & & 37\,fb          & 73\,fb         & 163\,fb       \\
    \bottomrule
  \end{tabular}
  
    \caption{The leading-order Higgs cross sections for the \higgsstrahlung and  $\PW\PW$-fusion
     processes  for $\mH=125\,\GeV$ at three centre-of-mass energies.
      The cross sections are calculated~\cite{bib:Whizard} including initial-state radiation and are shown for unpolarised electron/positron
      beams and assuming the baseline ILC polarisation of $P(\Pem,\Pep) = (-0.8,\,+0.3)$.
    \label{tab:higgsprod}}
\end{table*}

\subsection{The leptonic recoil mass measurement}

\begin{figure}[htb]
\centering
\includegraphics[width=6.0cm]{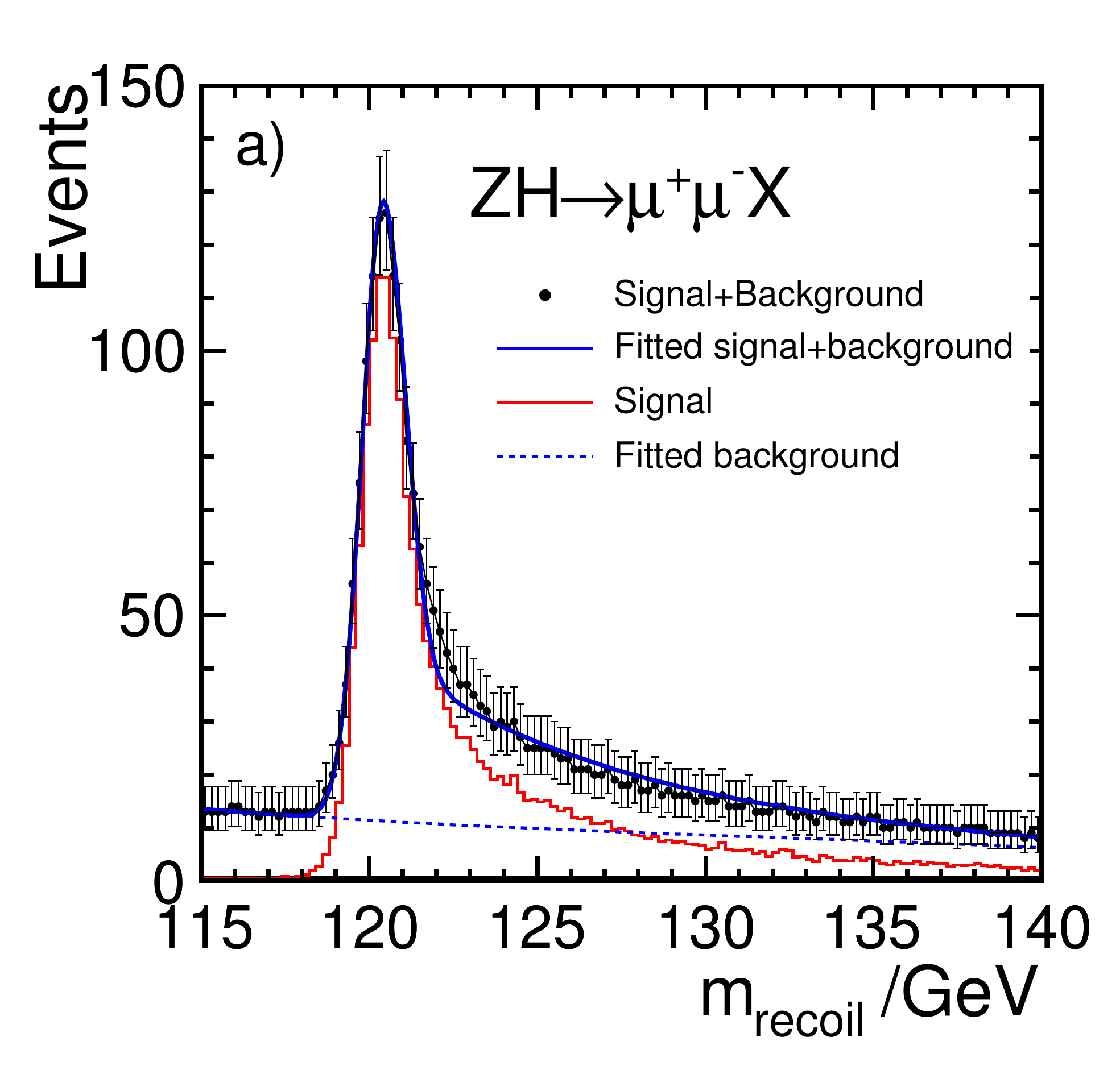}
\includegraphics[width=6.0cm]{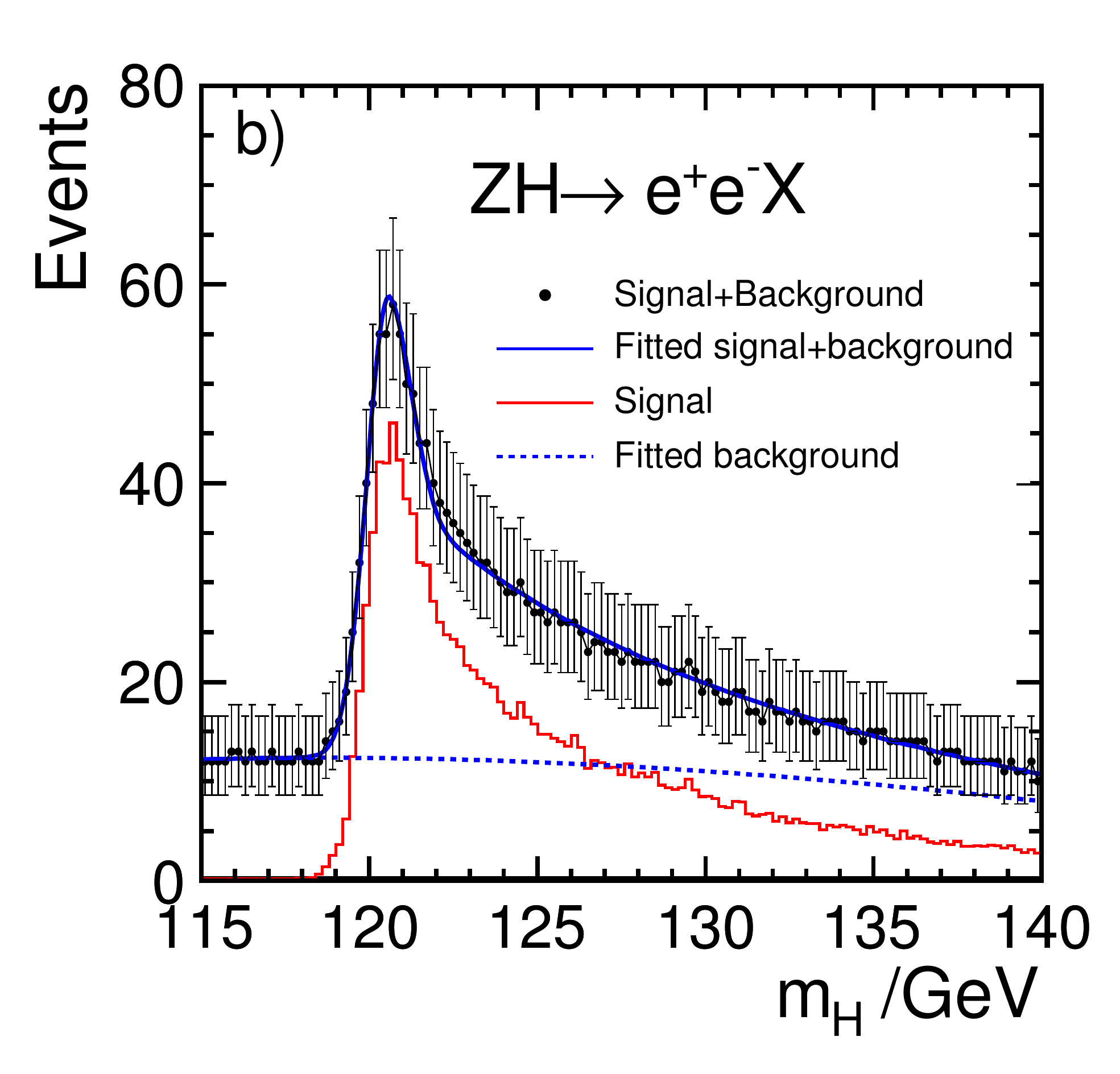}
\caption[Higgs recoil mass.]{The recoil mass distribution for the \higgsstrahlung process for
a) $\PZ\rightarrow\mpmm$ and b) $\PZ\rightarrow\epem$ at $\roots = 250\,\GeV$. The results are shown for $P(\Pe^-,\Pe^+) = (-80\,\%,+30\,\%)$ beam polarisation. 
Taken from~\cite{bib:ILD_LOI}.
\label{fig:leptonicRecoil}}
\end{figure}

The \higgsstrahlung process provides the opportunity to study the couplings of the Higgs boson in a
model-independent manner. This is unique to an electron-positron collider, where the
clean experimental environment and the relatively low SM cross sections for
background processes allow $\Pep\Pem\to\PH\PZ$ events to be selected based solely on the measurement of the 
four-momentum of the $\PZ$ boson, regardless of how the Higgs boson decays. The clearest topologies occur for $\PZ\to\epem$ and $\PZ\to\mpmm$ decays, 
which can be identified by first requiring that the measured di-lepton invariant mass $m_{\ell\ell}$ is consistent with $\mZ$. The 
four-momentum of the system recoiling against the $\PZ$ boson is obtained from 
$E_\mathrm{rec} = \roots - E_{\ell\ell}$ and $\vec{p}_\mathrm{rec} = -\vec{p}_{\ell\ell}$.  In $\Pep\Pem\to\PH\PZ$ events, 
the invariant mass of this recoiling system, $\mrec$, will peak at $m_{\PH}$. Fig.~\ref{fig:leptonicRecoil} shows the simulated 
recoil mass distribution in the ILD~\cite{bib:ILD_LOI} detector concept for 250\,\fbinv of ILC data at $\roots = 250\,\GeV$
with beam polarisation $P(\Pem,\Pep) = (-0.8,\,+0.3)$; by combining both $\PZ\to\mpmm$ and $\PZ\to\epem$
decays, $\sigma(\PH\PZ)$ can be measured to 2.6\,\%, leading to a determination of $\gHZZ$ with a precision of 1.3\,\%~\cite{bib:ILC_Physics_TDR, bib:ILD_LOI, bib:Li}. 

\subsection{Recoil mass measurement at different centre-of-mass energies}

The narrowness of the recoil mass peak is an important factor in determining the precision to which $\gHZZ$ can be measured. The recoil mass can be expressed as
\begin{align*}
     m_\text{rec}^2 &=  (  \roots - E_{\ell\ell} )^2 -\vec{p}_{\ell\ell}^2 
     =     s - 2\roots\,E_{\ell\ell} + E^2_{\ell\ell}  -\vec{p}_{\ell\ell}^2 \\ &=  s - 2\roots\, (E_{\ell1} + E_{\ell2}) + m^2_{\ell\ell} \,,
\end{align*}
where $\roots$ is the centre-of-mass energy and $E_{\ell1}$ and $E_{\ell2}$ are the energies of the two leptons. Since $m_{\ell\ell}$ will peak narrowly around $\mZ$, it can be seen 
that the width of the recoil mass peak scales with both $\roots$ and the lepton energy (or momentum) resolution. 
For high-momenta muons, where multiple scattering in the tracking chambers is relatively unimportant, the fractional momentum resolution $\sigma_p/p$ will scale 
approximately as the transverse momentum $\pT$, thus $\sigma_{E_\ell}$ will scale quadratically as $p\cdot\pT$.  
Consequently, in the range $\roots = 250 - 500\,\GeV$,  where the energy of the fermions from the $\PZ$ decay 
approximately scales as $\roots$, the width of the recoil mass distribution increases significantly with increasing centre-of-mass energy. For this reason, 
the leptonic recoil mass analysis leads to a higher precision on $\gHZZ$ for $\roots \sim 250\,\GeV$~\cite{bib:ILC_Physics_TDR}, where the $\sigma(\PH\PZ)$ is 
largest and the reconstructed recoil mass peak is relatively narrow. 
This has been one of the strongest arguments for the initial operation of the ILC at a relatively low centre-of-mass energy. 
This argument does not apply to the recoil mass measurement with hadronic $\PZ\to\qqbar$ decays since the recoil mass resolution 
depends less strongly on $\roots$ than for leptonic final states because the jet energy resolution for the linear collider detector concepts 
scales linearly with energy, $\sigma_E \sim 0.03 E$~\cite{bib:Thomson}. Although the hadronic recoil mass measurement has been considered previously~\cite{bib:Miyamoto}, 
this paper presents the first detailed study of its potential.

\section{Monte Carlo Samples, Detector Simulation and Event Reconstruction}

The CLIC results presented in this paper are based on detailed Monte Carlo (MC) simulation using: a full set of SM background processes; a detailed 
\geant~\cite{bib:Agostinelli, bib:Allison} simulation of the CLIC\_ILD detector concept~\cite{bib:CLIC_Physics_CDR}; 
and a full reconstruction of the simulated events.

\subsection{Monte Carlo event generation}

The simulated SM event samples were generated using the WHIZARD~1.95~\cite{bib:Whizard} program. The expected energy spectra for the 
CLIC beams, including the effects from beamstrahlung and the intrinsic machine energy spread, were used for the initial-state electrons and positrons.
The process of fragmentation and hadronisation of final-state quarks and gluons was simulated 
using PYTHIA 6.4~\cite{bib:PYTHIA} with a parameter set~\cite{bib:OPAL} that was tuned to OPAL $\epem$ data recorded at LEP. 
The decays of $\tau$ leptons were simulated using the TAUOLA package~\cite{bib:tauola}. The mass of the Higgs boson was taken to be $\mH = 126\,\GeV$ and the 
decays of the Higgs boson were simulated using PYTHIA with the branching fractions of \cite{bib:HiggsBR}. A dedicated sample of
$\epem\to\PH\PZ$ events with Higgs decays to ``invisible'' long-lived neutral particles was produced by artificially setting the Higgs boson 
lifetime to infinity. Because of the 0.5\,ns bunch spacing in the CLIC beams, the pile-up of beam-induced backgrounds from the \gghadrons process 
was included in the simulated event samples to ensure its effect on the event reconstruction was accounted for. 

\subsection{CLIC detector simulation and event reconstruction}

\label{sec:simreco}

The \geant-based Mokka~\cite{bib:Mokka} program was used to simulate the 
detector response of the CLIC\_ILD detector concept~\cite{bib:CLIC_Physics_CDR}. The \texttt{QGSP\_BERT} physics list was used to model the hadronic interactions of particles in the detectors. 
The hit digitisation and the event reconstruction were performed using the \marlin~\cite{bib:Marlin} software packages. 
Particle flow reconstruction was performed using PandoraPFA~\cite{bib:Thomson, bib:Marshall}.  
An algorithm, using the individual reconstructed particles, was used to identify and remove approximately 90\,\% of the out-of-time background 
due to pile-up from $\gghadrons$; here the {\sc{Loose}} particle flow object selection, described in~\cite{bib:CLIC_Physics_CDR}, was used.

Jet finding was performed using the \fastjet~\cite{bib:Fastjet} package.
Because of the presence of pile-up from  \gghadrons, the {\tt ee\_kt} (Durham) algorithm employed at LEP 
is not effective as it clusters particles from pile-up into the reconstructed jets. Instead, the hadron-collider inspired $k_t$ algorithm, 
with the distance parameter $R$ based on $\Delta\eta$ and $\Delta\phi$, was used with $R=\pi/2$. This
algorithm allows particles to be clustered into ``beam jets'', aligned with the beam axis, in addition to jets seeded by high-momentum particles.
Background from the pile-up of $\gghadrons$ can, to a large degree, be removed by ignoring particles in the ``beam jets'', 
largely mitigating the impact of the beam background.

The hadronic recoil mass study, presented in this paper, covers a wide range of $\PH\PZ(\PZ\to\qqbar)$ final-state topologies ranging from two jets where
Higgs decays to long-lived neutral particles, $\epem\to\PH\PZ\to(\text{invis.})(\qqbar)$, to six-jet toplogies from, for example, 
$\epem\to\PH\PZ\to(\PW\PW^*\to\qqbar\qqbar)(\qqbar)$. For this reason, each reconstructed event is clustered into two-, three-, four-, five- and six-jet
topologies, with ``$y$-cut'' variables used to indicate the underlying physical topology.  
For example, if an event is forced into a three-jet topology, $y_{34}$ is the $k_t$ value at which the event would be reconstructed as four jets and
$y_{23}$ is the $k_t$ value at which the event would be reconstructed as two jets. 

\subsection{ILC detector simulation and  event reconstruction}

The event generation and reconstruction for the ILC studies, presented in section~\ref{sec:ilc}, follows closely that described above. The main
differences are: i) the ILC beam spectrum, where the effects of beamstrahlung are less pronounced, 
was used, ii) the detector simulation used the ILD detector concept for the ILC, rather than the CLIC\_ILD model adapted for CLIC; and iii) 
the much longer bunch spacing at the ILC means that only in-time background from $\gghadrons$ needs to be included.   

\section{Hadronic Recoil Mass Measurement at CLIC}

\label{sec:clicanalysis}

In the process $\epem\to\PH\PZ$ it is possible to cleanly identify $\PZ\to\epem$ and $\PZ\to\mpmm$ decays 
regardless of the $\PH$ decay mode. Consequently, the selection efficiency is almost independent of 
the nature of the $\PH$ decay. For $\PZ\to\qqbar$ decays, the selection efficiency will depend more strongly on the 
Higgs decay mode. For example, in $(\PH\to\PQb\PAQb)(\PZ\to\qqbar)$ events, the reconstruction 
of the $\PZ$ boson is complicated by mis-associations of particles to jets and by the three-fold ambiguity 
in associating four jets to the $\PZ$ and $\PH$. These ambiguities will increase with the number of jets in 
the final state. For this reason, it is much more difficult to construct an event selection, based only on the reconstructed 
candidate $\PZ\to\qqbar$ decay, with a selection efficiency that is independent of the Higgs decay mode. 
Nevertheless it is possible to minimise this dependence. The strategy adopted here is to: 
i) separate all simulated events into candidates for Higgs decays to ``invisible'' long-lived neutral particles and decays to visible final states; 
ii) identify the di-jet system that is the best candidate for the $\PZ\to\qqbar$ decay; 
iii) reject events consistent with a number of clear background topologies using the information from the whole event; 
iv) identify $\PH\PZ(\PZ\to\qqbar)$ events solely based on the properties from the candidate $\PZ\to\qqbar$ decay, first for the candidate visible
Higgs decays and then for the candidate invisible Higgs decays; and v) combine the results into a single measurement of $\sigma(\PH\PZ)$.

\subsection{Separation into candidate visible and invisible Higgs decay samples}

Hadronic events are selected by forcing each event into a two-jet topology and requiring at least three charged 
particles in each jet. The surviving events are then divided into candidates for either visible $\PH$ decays or invisible $\PH$ decays, in both cases 
produced in association with a $\PZ\to\qqbar$. 
Events are categorised as potential invisible $\PH$ decays on the basis of the $y$-cut values in the $k_t$ jet-finding algorithm. For invisible
$\PH$ decays, only the $\PZ\to\qqbar$ system is visible in the detector, typically resulting in a two-jet topology (with the possibility that 
QCD radiation can increase the number of reconstructed jets). Consequently, invisible $\PH$ decays will have small values of $y_{23}$ and $y_{34}$, 
the variables respectively representing the $k_t$ value at which an event transitions from two to three jets and from three to four jets,
as indicated in Fig.~\ref{fig:y23_y34}. Events are categorised as candidate invisible $\PH$ decays if 
$-\log_{10}(y_{23})> 2.0$ and $-\log_{10}(y_{34})> 3.0$. 
Due to gluon radiation in the parton shower, only 74\,\% of the simulated $\PH\PZ\, (\PZ\rightarrow\qq)$ events with 
invisible $\PH$ decays are placed in this two-jet topology candidate invisible $\PH$ decay sample. 
To improve the efficiency for correctly categorising SM Higgs decays with low-energy leptons, for example $\PH\to\PW\PW^*\to\PGt\PGn\PGt\PGn$, 
events with $-\log_{10}(y_{23})<2.5$ and $-\log_{10}(y_{34})< 3.5$ are forced into three jets and are excluded from the invisible Higgs decay sample 
if the lowest-energy jet has fewer than four reconstructed tracks or contains an identified $\Pe^\pm/\PGm^\pm$ with energy $E>5\,\GeV$. 
Only 2.2\,\% of simulated $\PH\PZ$ events with SM Higgs decays end up in the candidate invisible Higgs sample.    

{\begin{figure*}[hbt]
\centering
      \includegraphics[width=0.84\textwidth]{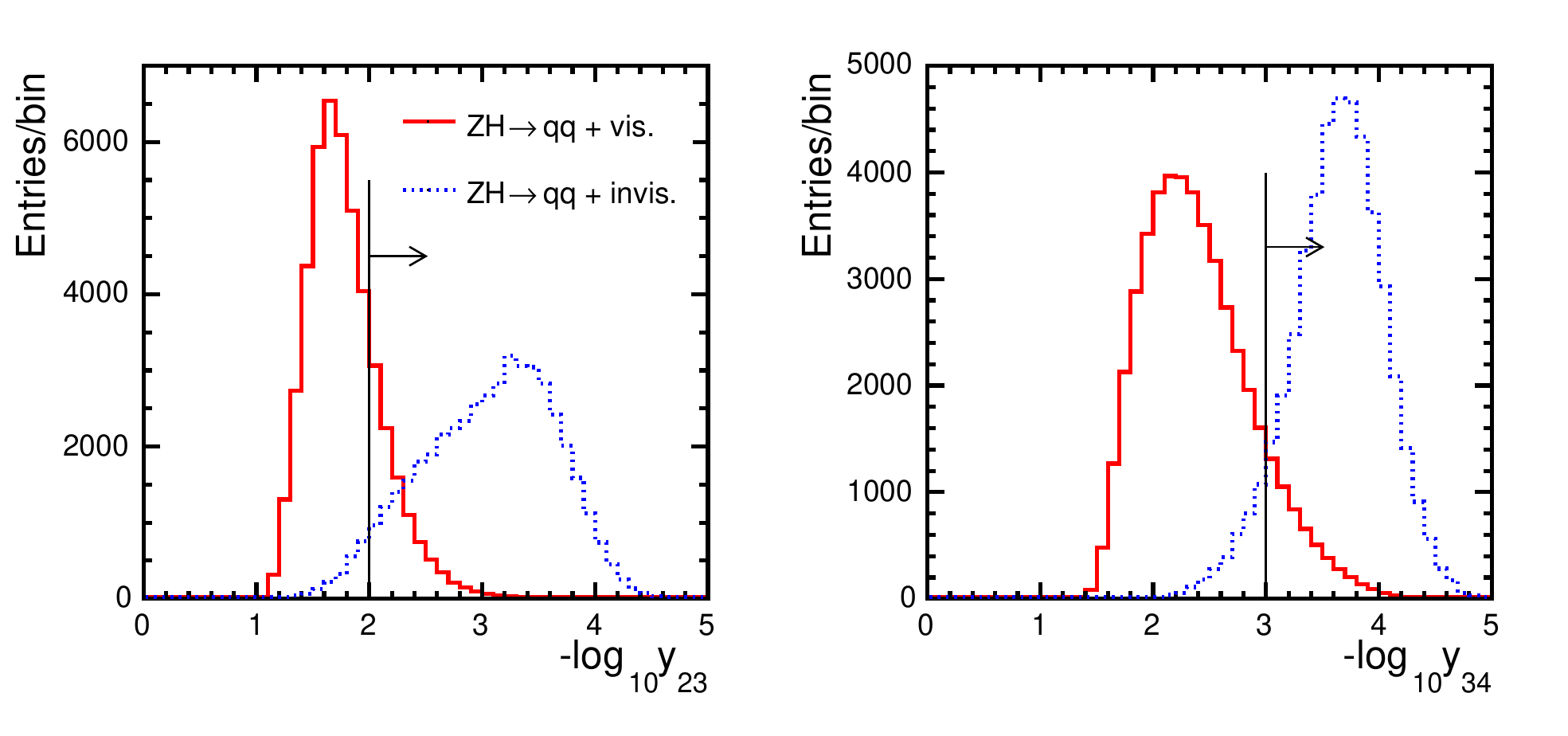}   
\caption{\label{fig:y23_y34} The distributions of $-\log_{10}(y_{23})$ and $-\log_{10}(y_{34})$ for simulated $\PH\PZ\,(\PZ\to\qqbar)$ events
      for visible and invisible Higgs decays at $\roots = 350\,\GeV$. The distributions are normalised to an integrated luminosity of 500\,\fbinv. 
      The distribution for the invisible $\PH$ decays assumes a 100\,\% branching fraction into invisible decay modes. The vertical lines with arrows 
      indicate the cut values used in this analysis.}
\end{figure*}}

\subsection{Recoil mass reconstruction}

For each candidate $\PH\PZ(\PZ\to\qqbar)$ event, the recoil mass is calculated from
$\mrec^2 =  (  \roots - E_{\qqbar} )^2 -\vec{p}_{\qqbar}^2$, where $E_{\qqbar}$ and $\vec{p}_{\qqbar}$ are the summed energy and
momentum of the di-jet system from the identified candidate $\PZ\to\qqbar$ decay.  In the case of the candidate invisible Higgs decay sample, 
the two jets are assumed to be from $\PZ\to\qqbar$. The resulting recoil mass distribution for candidate invisible Higgs
decays, which is strongly peaked around $\mrec \sim \mH$, is shown in Fig.~\ref{fig:mrec}a. In the case of the candidate visible Higgs decay 
sample, the situation is more complicated as this sample encompasses many different $\PH\PZ$ event topologies. 
For example, $\PH\to\PQb\PAQb$ decays will result in a four-quark $\PH\PZ$ final state, usually yielding four jets,
whereas, $\PH\to\PW\PW^*\to\qqbar\Pl\PGn$  and  $\PH\to\PW\PW^*\to\qqbar\qqbar$ decays will respectively usually yield five- and six-jet final states. 
In all cases gluon radiation in the parton shower can increase the reconstructed jet multiplicity relative to the tree-level expectation. 

In order to achieve the desired (near) model independence of the analysis, it is necessary to have a similar quality of recoil mass reconstruction for all 
Higgs boson visible decay modes. This hinges on the correct identification and reconstruction of the $\PZ\to\qqbar$ di-jet system. 
The first stage is to force events in the candidate visible Higgs decay sample into a four-jet topology. From the three possible di-jet combinations, 
the di-jet system with invariant mass $\mqq$ closest to $\mZ$ is identified as the candidate $\PZ\to\qqbar$ decay and its energy and 
momentum are used to calculate the recoil mass $\mrec$. In selecting the candidate $\PZ$ decay, only jets containing more than three charged particles are considered. 
To improve the reconstruction of higher-jet-multiplicity final states, such as  $\PH\to\PW\PW^*\to\qqbar\qqbar$, each event is also forced into five jets 
and the di-jet system with mass closest to $\mZ$ is again identified as the candidate $\PZ\to\qqbar$ decay. The five-jet topology is only used if  $-\log_{10}(y_{45})>3.5$ and 
both $\mqq$ and $\mrec$ are respectively closer to $\mZ$ and $\mH$ than the corresponding values from the four-jet reconstruction.
Even in the genuine six-parton topology $\PH\PZ\to(\PW\PW^*)\qqbar\to(\qqbar\qqbar)\qqbar$ only 13\,\% of events are reconstructed as five jets, for the remainder, the four-jet reconstruction is preferred. However,  
provided the jets from the $\PZ\to\qqbar$ decay are correctly identified, there is no need to correctly reconstruct the recoiling system as only the 
properties of the  $\PZ\to\qqbar$ decay  are used in the subsequent analysis. For this reason, allowing the possibility of reconstructing events as six jets 
was found not to improve the overall recoil mass reconstruction.  Fig.~\ref{fig:mrec}b  shows the resulting recoil mass distribution for
simulated $\PH\PZ$ events with $\PH\to\PQb\PAQb$, $\PH\to\PW\PW^*\to\qqbar\qqbar$ and $\PH\to\tptm$. Despite the very different final states, 
similar recoil mass distributions are obtained.  

{\begin{figure*}[bth]
  \centering
      \includegraphics[width=0.88\textwidth]{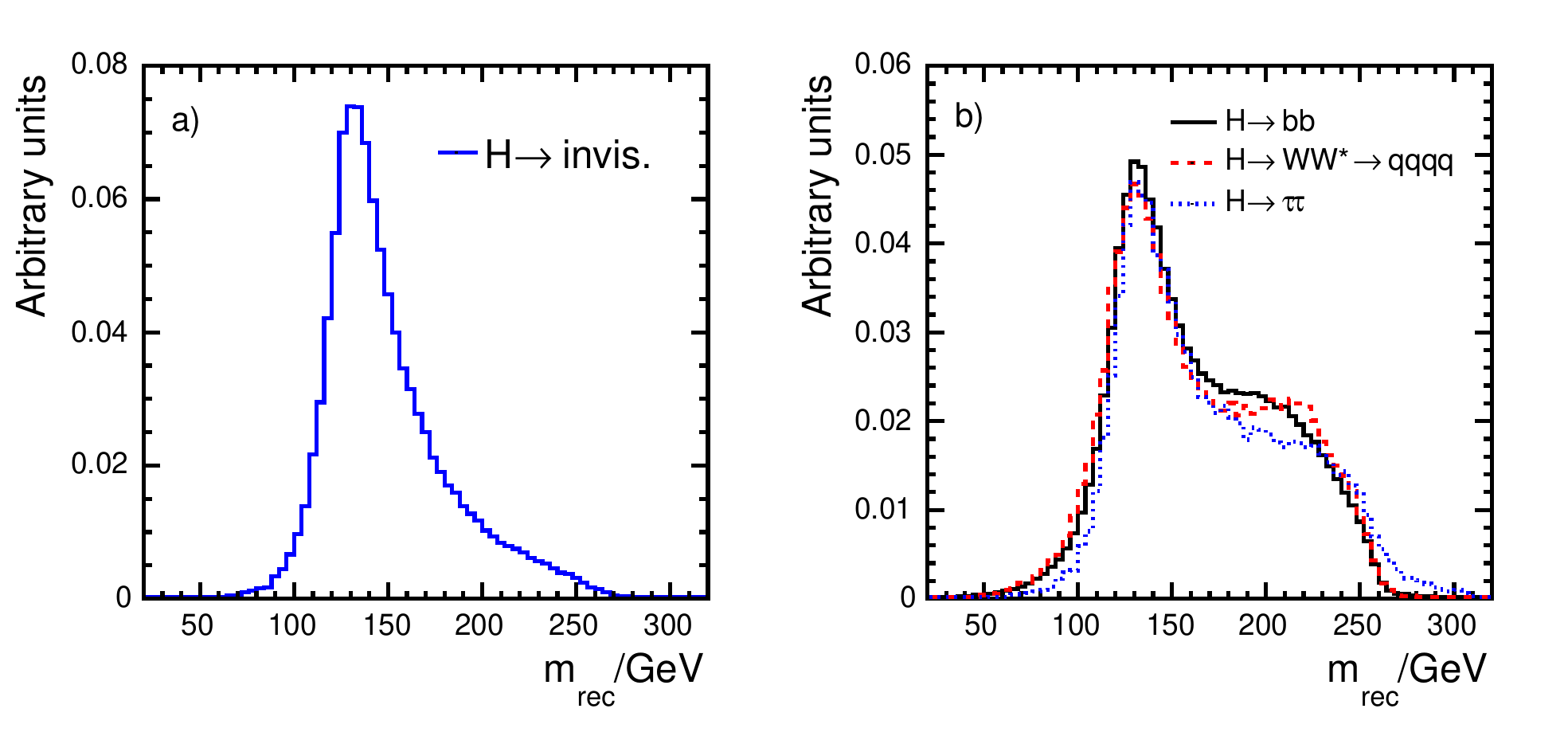} 
      \caption{\label{fig:mrec} a) The reconstructed hadronic recoil mass distribution for the candidate $\PH\PZ$ events with $\PZ\to\qqbar$ and 
         $\PH\to\text{invis.}$ b) The reconstructed hadronic recoil mass distributions for 
         candidate $\PH\PZ(\PZ\to\qqbar)$ and either $\PH\to\PQb\PAQb$, $\PH\to\PW\PW^*\to\qqbar\qqbar$ or $\PH\to\tptm$. In each case the distributions are normalised to unit area. An underflow (not shown) contains the small fraction of events where no good $\PZ\to\qqbar$ candidate is identified.}
\end{figure*}}

\subsection{Preselection}

\label{sec:presel}

After dividing all events into either candidates for visible or invisible Higgs decays and having identified the two jets forming the candidate $\PZ\to\qqbar$ system, 
preselection cuts are applied to reduce backgrounds from larger cross section SM processes such as $\epem\to\PQq\PAQq$ 
and \mbox{$\epem\to\PQq\PAQq\PQq\PAQq$}. Cuts are based on the invariant mass of the $\PZ\to\qqbar$ candidate, $\mqq$,
and corresponding recoil mass, $\mrec$.  In addition, the invariant mass of all the visible particles not originating from the 
candidate $\PZ\to\qqbar$ decay,  $\mqqp$, is calculated. It is important to note that $\mqqp$ is only used to reject specific background topologies in the
preselection and is not used in the main selection; in $\PH\PZ$ events $\mqqp$ will depend strongly on the Higgs decay mode. 
The preselection cuts (most of which are common to the visible and invisible Higgs selections) are:
\begin{itemize}
   \item  the event must be broadly consistent with being $\PH\PZ$, $70\,\GeV < \mqq < 110\,\GeV$ and  $80\,\GeV < \mrec < 200\,\GeV$. 
   \item  background from $\epem\to\qqbar$ is suppressed 
            by removing events with net transverse momentum $\pT < 20 \,\GeV$ and
            $-\log_{10}(y_{34})>2.5$, indicating a final-state system consisting of fewer than four primary particles.
   \item events in the invisible Higgs decay sample are rejected if $|\cos\thetamis|>0.7$, where
            $\thetamis$ is the polar angle of the missing momentum vector,  almost completely eliminating the
            contribution from $\epem\to\qqbar$ with unobserved initial-state radiation (ISR).  
   \item events in the invisible Higgs decay sample are rejected if there is an isolated identified $\Pe^\pm/\PGm^\pm$ with energy $E_\Pl>10\,$GeV,
            suppressing background from $\PW\PW\to\PQq\PAQq\Pl\PGn$. 
    \item  background from $\epem\to\qqbar$ with unobserved ISR, including radiative return to the $\PZ$ resonance, 
            is suppressed by rejecting events with net transverse momentum $\pT < 20 \,\GeV$ and $|\cos\thetamis|>0.9$.       
   \item the background from $\epem\to\PW\PW\rightarrow\qqbar\qqbar$ is suppressed by
            forcing events into four jets and selecting the di-jet pair with the mass $\mqq$ closest to $\mW$. Events are rejected if
            $\pT < 20 \,\GeV$ and $65\,\GeV< \mqq < 100\,\GeV$ and $65\,\GeV< \mqqp < 100\,\GeV$, where 
            $\mqqp$ is the measured invariant mass of the second di-jet pair.
  \item the background from $\epem\to\PZ\PZ\rightarrow\qqbar\qqbar$ is suppressed in a similar manner. Events
            are forced into four jets and the di-jet pair with the $\mqq$ closest to $\mZ$ is identified. Events are rejected if
            $\pT < 20 \,\GeV$ and $70\,\GeV< \mqq < 105\,\GeV$ and $70\,\GeV< \mqqp < 105\,\GeV$,  where 
            $\mqqp$ is the measured  mass of the second di-jet pair.
\end{itemize}

 The effects of the preselection cuts are summarised in Tab.~\ref{tab:presel}. The events passing the preselection cuts are put forward as
candidate $\PH\PZ(\PZ\to\qqbar)$ events with either: i) visible $\PH$ decay products;  or ii) invisible $\PH$ decay products, 
depending on whether the event was consistent with a two-jet topology or not. The first two cuts listed above result in the largest loss
of signal efficiency for the visible Higgs decay selection. The $\PQq\PAQq\Pl\PGn$ background in the visible Higgs decay 
preselection could have been significantly reduced by rejecting events with visible high-energy isolated leptons, but this would have introduced a bias against $\PH$ 
decays with leptons in the final state.

{\begin{table}[htb]
  \centering
   \begin{tabular}{lrrrrr}
    \toprule 
         Final state                                & \tabt{$\sigma/\text{fb}$} & \tabt{$\varepsilon^\text{vis.}_\text{presel}$} & \tabt{$\varepsilon^\text{invis.}_\text{presel}$}   & \tabt{$N^\text{vis.}_\text{presel}$} & \tabt{$N^\text{invis.}_\text{presel}$}    \\ \midrule
    $\PQq\PAQq$                                            & 25180   &     0.4\,\%       & $-$\phantom{\,\%}    & 54570     & 0 \\
      $\PQq\PAQq\Pl\PGn$                              &  5914      &   11.2\,\%       & 0.9\,\%                      & 326420 & 26060 \\   
      $\PQq\PAQq\PQq\PAQq$                        & 5847       &     3.8\,\%       & $-$\phantom{\,\%}    & 110520 & 0       \\    
     $\PQq\PAQq\Pl\Pl$                                   &  1704    &      1.5\,\%       &  $-$\phantom{\,\%}    & 13260   & 60     \\ 
     $\PQq\PAQq\PGn\PAGn$                         &  325       &      0.6\,\%       &  14.8\,\%                    & 1050   &  24180  \\    
     $\PH\PGne\PAGne$                                 &  52       &      2.5\,\%      &  5.6\,\%                      & 640         &  1430 \\
  $\PH\PZ$; $\PZ\to\PQq\PAQq$                  &   93      &     42.0\,\%      &   0.2\,\%                   & 19630       &  80 \\
  \midrule
  $\PH\to\text{invis.}$\, (100\,\%) \!\!\!\!\!\!\!\!\!\!\!\!\!                   &   93     &       0.6\,\%      &   48.6\,\%                 &  300          & 22710\\                                                                                                                                                                           
    \bottomrule
  \end{tabular}
   \caption{Summary of the effects of the preselection cuts for the the visible and invisible recoil mass analyses. The
     efficiencies $\varepsilon$ include the effects of the preselection cuts and the division into the candidate visible and invisible Higgs decay samples. 
     The expected numbers of events passing the preselection cuts 
     correspond to an integrated luminosity of $500\,\text{fb}^{-1}$ at CLIC, assuming unpolarised beams at $\roots=350\,\GeV$.  The numbers shown for the invisible Higgs decay modes correspond to a 100\,\% branching ratio. 
 \label{tab:presel}}
\end{table}}
 
 \subsection{Selection of HZ$\to$q$\overline{\text{q}}$ with visible Higgs decays} 

After preselection, the main backgrounds in the visible Higgs decay analysis arise from
$\epem\to\PQq\PAQq\Pl\PGn$ and $\epem\to\PQq\PAQq\PQq\PAQq$, dominated by 
$\PW\PW$, single-$\PW$ ($\PW\Pe\PGne$) and $\PZ\PZ$ processes. The event selection is based entirely 
on the reconstructed candidate $\PZ\to\qqbar$ decay in the event. The properties of the 
remainder of the event (or the event as a whole) are not used as their inclusion would
break the desired model independence of the selection. For example, the background from 
$\epem\to\PQq\PAQq\Pl\PGn$ can be significantly reduced by placing a lower bound on the total 
visible energy in the event, however such a cut would bias the selection against Higgs decays with missing
energy, such as ${\PH\to\tptm}$. 

The event selection uses a relative likelihood approach with discriminant variables based on the properties of
candidate $\PZ\to\qqbar$ decay. Two event categories are considered: a) the $\epem\to\PH\PZ\to\PH\qqbar$ signal; and b) 
all non-Higgs background processes. The relative likelihood for an event being classified as signal is defined as
\begin{equation*}
       {\cal{L}}    = \frac{L_\text{signal} }{L_\text{signal} + L_\text{back}} \,,
\end{equation*}  
where the individual absolute likelihood $L_j$ for the event class $j$ (signal or background) is formed from normalised probability distributions 
$P^j_i(x_i)$ of the discriminant variables $x_i$ for that event class $j$:
\begin{equation*}
    L_j = \sigma^j_\text{presel} \times \prod_i^{N} P^j_i(x_i) \,,
\end{equation*}  
where $\sigma^j_\text{presel}$ is the cross section after preselection for event class $j$. 

The discriminant variables used in the likelihood selection, all of which are based on the candidate $\PZ\to\qqbar$ decay, are:  
{i)} the two-dimensional distribution of $\mqq$ and $\mrec$; {ii)} the polar angle of the $\PZ$ candidate, $|\cos\theta_{\PZ}|$; 
and {iii)} the modulus of the angle of the jets from the $\PZ\to\qq$ decay in $\PZ$ rest frame, relative to its laboratory frame direction of motion, $\cosQ$. 
The two-dimensional distributions of $\mqq$ and $\mrec$, are shown separately for the signal and background 
in Fig.~\ref{fig:visMass}. As expected, the $\PH\PZ$ signal events peak around $\mqq\approx\mZ$ and $\mrec\approx\mH$. 
The anti-correlation between $\mrec$ and $\mqq$ is expected; when the 
reconstructed jet energies are higher than the true energies, the reconstructed value of $\mqq$ will be higher than $\mZ$ and 
$\mrec$ will be lower than $\mH$ due to the  $-2\!\roots\,E_{\PZ}$ term in the expression for 
the recoil mass, $\mrec^2= s - 2\!\roots\,E_{\PZ} + \mqq^2\,$.  
The broad peaked structure in the background distribution at lower values of $\mqq$ arises from $\PQq\PAQq\Pl\PGn$ events (which have been 
forced into a four- or five-jet topology). The use of the two-dimensional distribution of $\mqq$ versus $\mrec$ in the likelihood accounts for
the associated correlations.  

{\begin{figure}[htb]
\centering
      \includegraphics[width=0.49\textwidth]{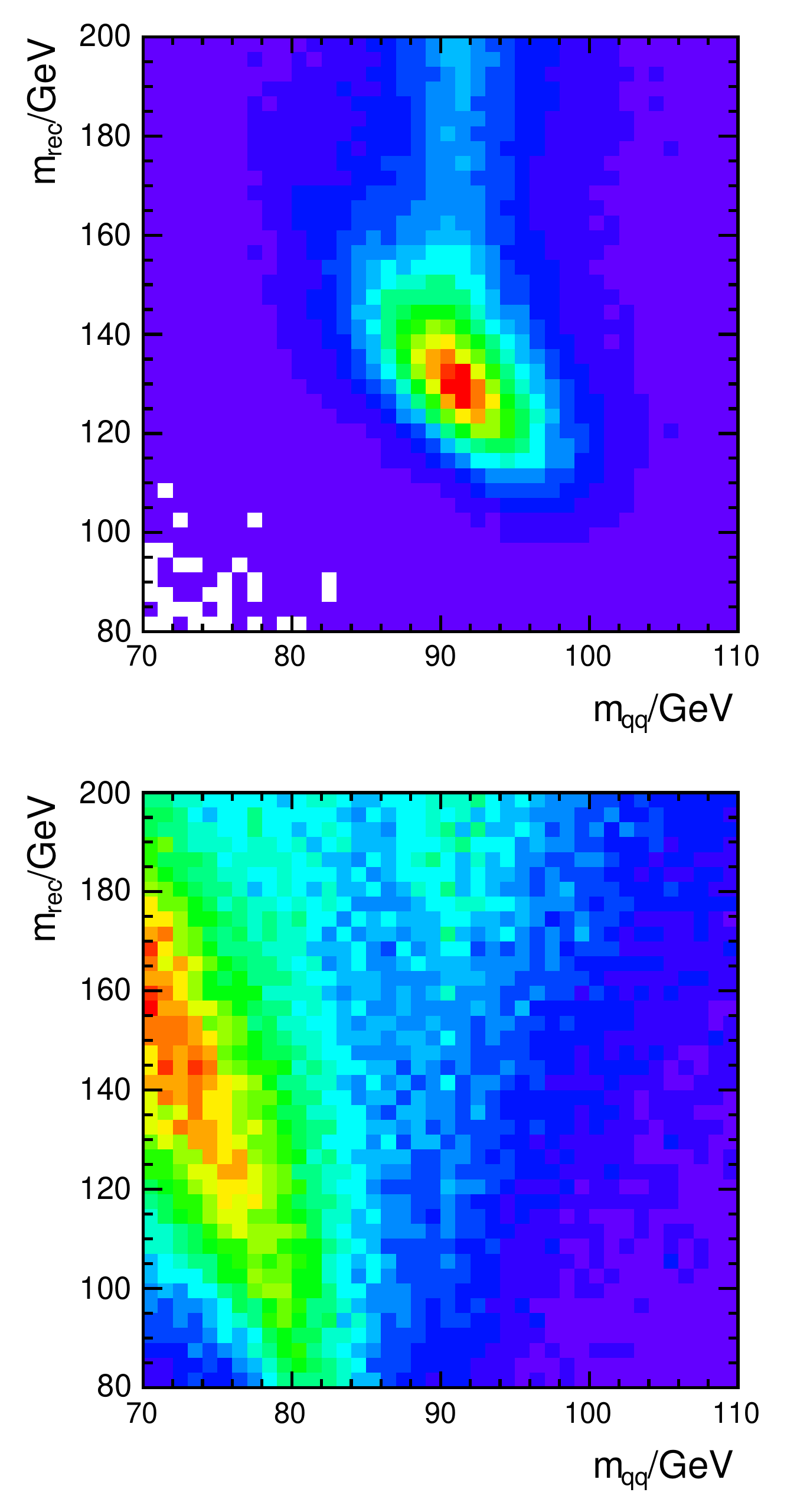} 
      \caption{\label{fig:visMass} The distribution of the reconstructed $\PZ\to\qq$ mass, $\mqq$, versus the hadronic recoil mass, $\mrec$, for $\PH\PZ(\PZ\to\qqbar)$ events (top) and for all background processes (bottom). In both cases the plots show all events passing the visible Higgs preselection for CLIC operating at $\roots=350\,\GeV$.}
\end{figure}}

The two angular variables used in the likelihood selection are shown in Fig.~\ref{fig:visSelAngles}. The discriminating 
power arises from the fact that the Higgs boson is a scalar particle, and the angular distributions in $\PH\PZ$ production are 
different from those in the dominant backgrounds which mostly arise from the production of two vector particles.

{\begin{figure*}[htb]
\centering
      \includegraphics[width=0.9\textwidth]{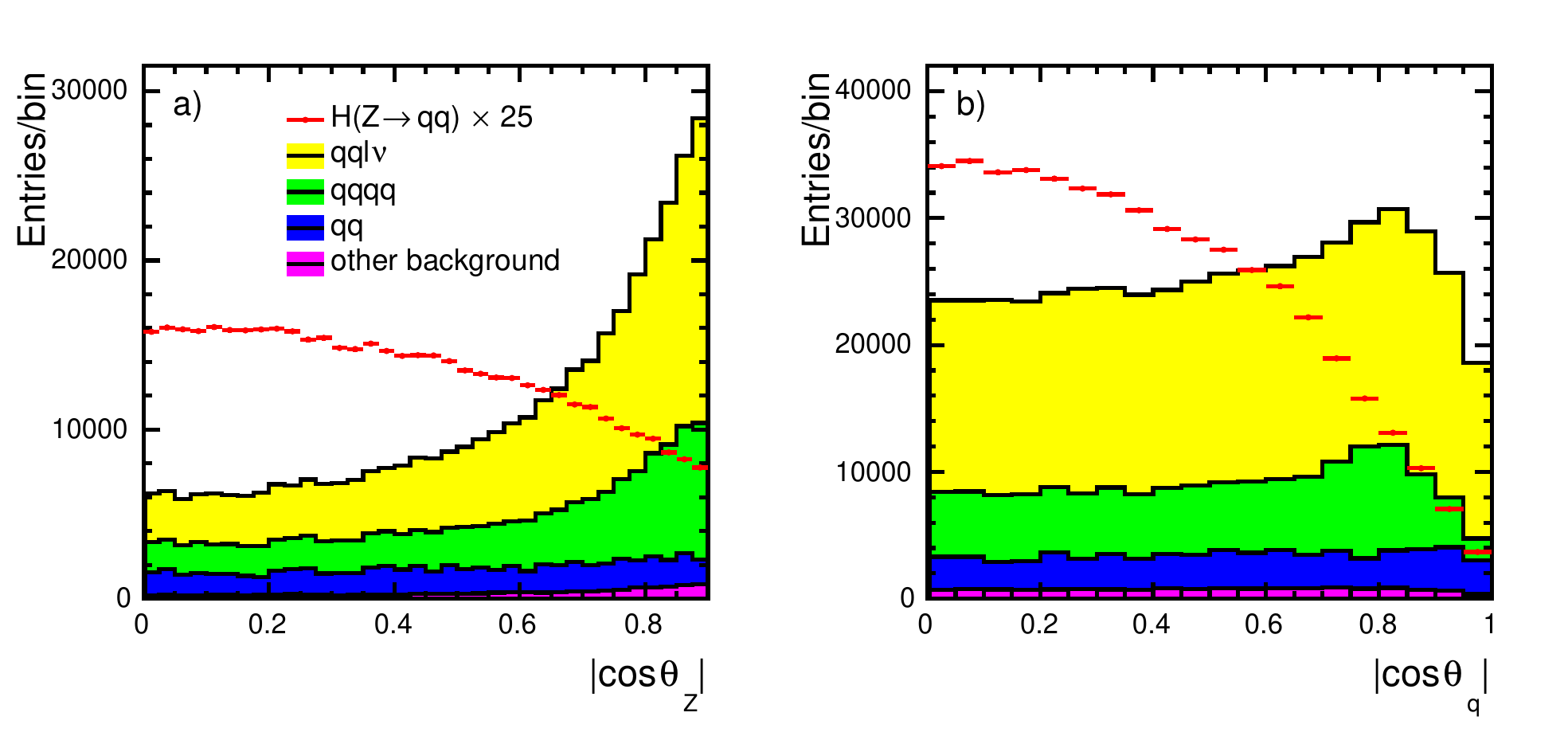}
      \caption{\label{fig:visSelAngles} a) The polar angle of the reconstructed $\PZ$ candidates, $|\cos\theta_{\PZ}|$, for both signal and background events for CLIC operating at $\roots=350\,\GeV$, and
       b) the modulus of the angle of the jets from the $\PZ\to\qq$ decay relative to the $\PZ$ direction after boosting into 
       its rest frame, $\cosQ$.  The signal and background distributions are normalised to 500\,\fbinv, but the $\PH\PZ(\PZ\to\qqbar)$ signal has been scaled by a factor of 25 to improve its
      visibility.}
\end{figure*}}

The resulting relative likelihood distribution is shown in Fig.~\ref{fig:visLike}. Despite the fact that the signal-to-background ratio in the preselected event
sample is approximately 1:25, the likelihood selection provides good separation. The statistical precision on the cross section for $\PH\PZ$ production
(where the $\PZ$ decays hadronically and the $\PH$ has SM branching fractions) is maximised with a likelihood cut of ${\cal{L}}>0.65$. The resulting 
efficiencies and the expected numbers of selected events for an integrated luminosity of 500\,$\fbinv$ are shown in Tab.~\ref{tab:visSel}. The corresponding statistical 
uncertainty on the 
production cross section is $\pm1.9\,\%$. The precision can be improved by extracting the number of signal events by performing a maximum-likelihood fit to the shape of the simulated likelihood distribution by varying the normalisations of the signal and background components, yielding a statistical error of
\begin{align*}  
        \Delta \sigma_\text{vis.} &= \pm 1.7\,\% \,.
\end{align*}

{\begin{figure}[htb]
\centering
    \includegraphics[width=0.45\textwidth]{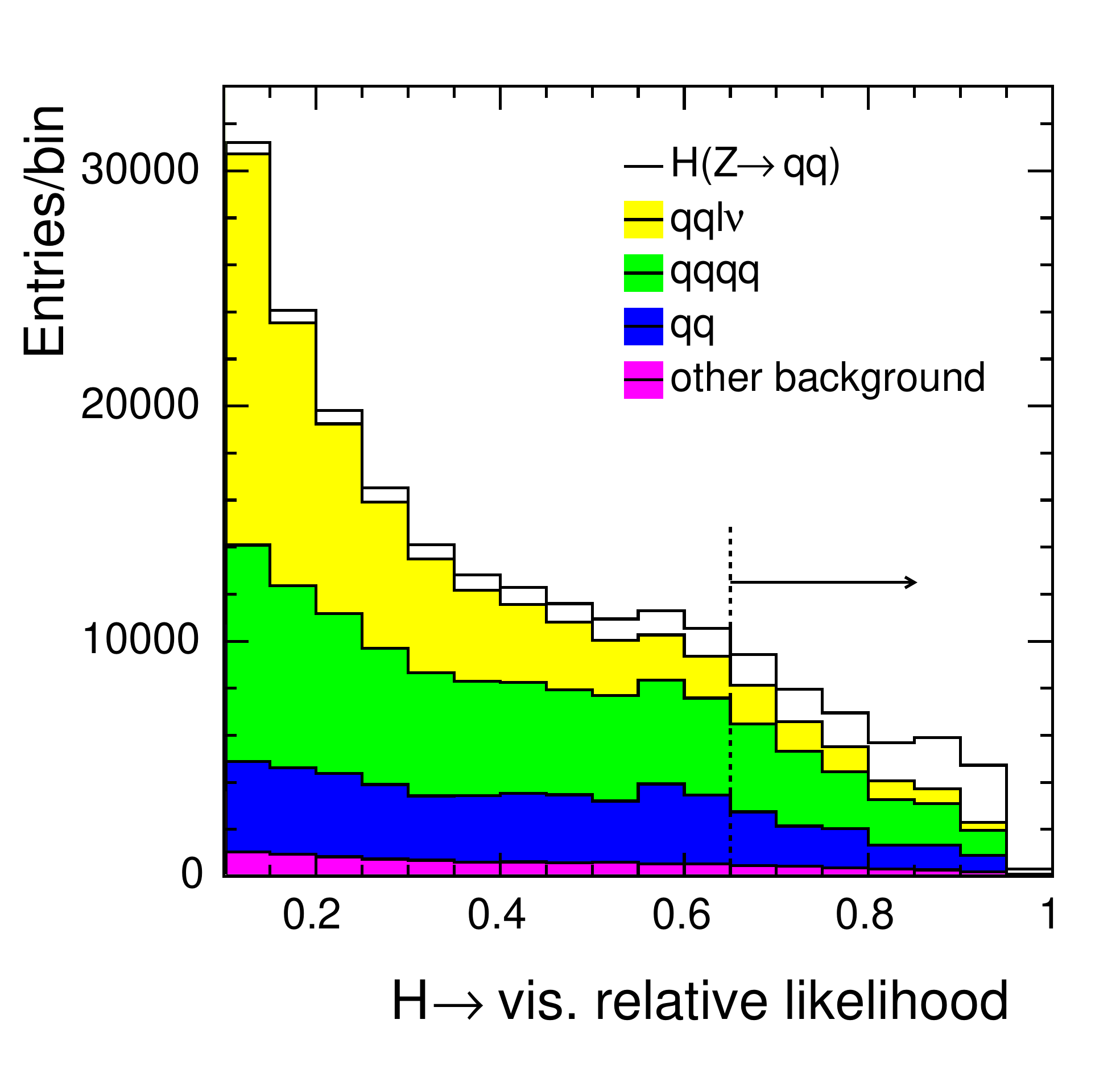} 
      \caption{\label{fig:visLike} The resulting likelihood distribution for the hadronic recoil mass analysis. The distributions correspond to 
         500\,\fbinv of CLIC operation at $\roots=350\,\GeV$ with unpolarised electron and positron beams. 
         The optimal likelihood cut at ${\cal{L}}=0.65$ is indicated by the arrow.}
\end{figure}}

{\begin{table}[bht]
  \centering
   \begin{tabular}{lrrrr}
    \toprule 
            Process                                & \tabt{$\sigma/\text{fb}$} & \tabt{$\varepsilon_\text{presel}$} & \tabt{$\varepsilon^\text{vis.}_{{\cal{L}}>0.65}$} & \tabt{$N_{{\cal{L}}>0.65}$} \\ \midrule
    $\PQq\PAQq$                                            & 25180   &     0.4\,\%       &  0.07\,\%                     & 8525  \\
      $\PQq\PAQq\Pl\PGn$                              &  5914      &   11.2\,\%       & 0.20\,\%                   & 5767 \\   
      $\PQq\PAQq\PQq\PAQq$                        & 5847       &     3.8\,\%       & 0.49\,\%                   & 14142    \\    
     $\PQq\PAQq\Pl\Pl$                                   &  1704    &      1.5\,\%       &  0.22\,\%                   & 1961     \\ 
     $\PQq\PAQq\PGn\PAGn$                         &  325       &      0.6\,\%       &  0.04\,\%                    & 60  \\    
     $\PH\PGne\PAGne$                                 &  52       &      2.5\,\%      &   0.23\,\%                      & 60   \\
  $\PH\PZ$; $\PZ\to\PQq\PAQq$                  &   93      &     42.0\,\%      &   22.6\,\%                   & 10568 \\
  \midrule
  $\PH\to\text{invis.}$ \, (100\,\%)                                       &   [93]     &       0.6\,\%      &   0.04\,\%                 &  20  \\                                                                                                                                                                           
    \bottomrule
  \end{tabular}
    \caption{Summary of the CLIC $(\PH\rightarrow\text{vis.})(\PZ\rightarrow\qq)$ event selection at $\roots=350\,\GeV$, giving the cross sections, preselection efficiency, overall selection   efficiency for a likelihood cut of  ${\cal{L}}>0.65$ and the expected numbers of events passing the 
  event selection for an integrated luminosity of $500\,\text{fb}^{-1}$ assuming unpolarised electron and positron beams. The numbers shown for the invisible Higgs decay modes correspond to a 100\,\% branching ratio. 
 \label{tab:visSel}}
\end{table}}

\subsection{Selection of $HZ\to q\overline{q}$ with invisible Higgs decays} 
 
The main backgrounds after preselection for the invisible Higgs decay selection arise from
$\epem\to\PQq\PAQq\Pl\PGn$ and $\epem\to\PQq\PAQq\PGn\PAGn$, which are dominated 
respectively by single-$\PW$ ($\PW\Pe\PGne$) and $\PZ\PZ$ processes. A relative likelihood selection is
used to separate the $(\PH\to\text{invis.})(\PZ\to\qqbar$) signal from the non-Higgs background. 
The discriminant variables employed are the same as those used for the visible Higgs decay likelihood function, namely
the two-dimensional distribution of $\mqq$ versus $\mrec$,  $|\cos\theta_{\PZ}|$ and $\cosQ$.        
The most powerful of these variables is the recoil mass itself, shown in Fig.~\ref{fig:invisSel}a, where the signal 
is plotted for the artificial case of $\BR(\PH\to\text{invis.}) = 100\,\%$.  The resulting relative likelihood 
distribution for signal and background is shown in 
Fig.~\ref{fig:invisSel}b, where good separation between signal and background is achieved.   

{\begin{figure*}[htb]
\begin{centering}
      \includegraphics[width=0.9\textwidth]{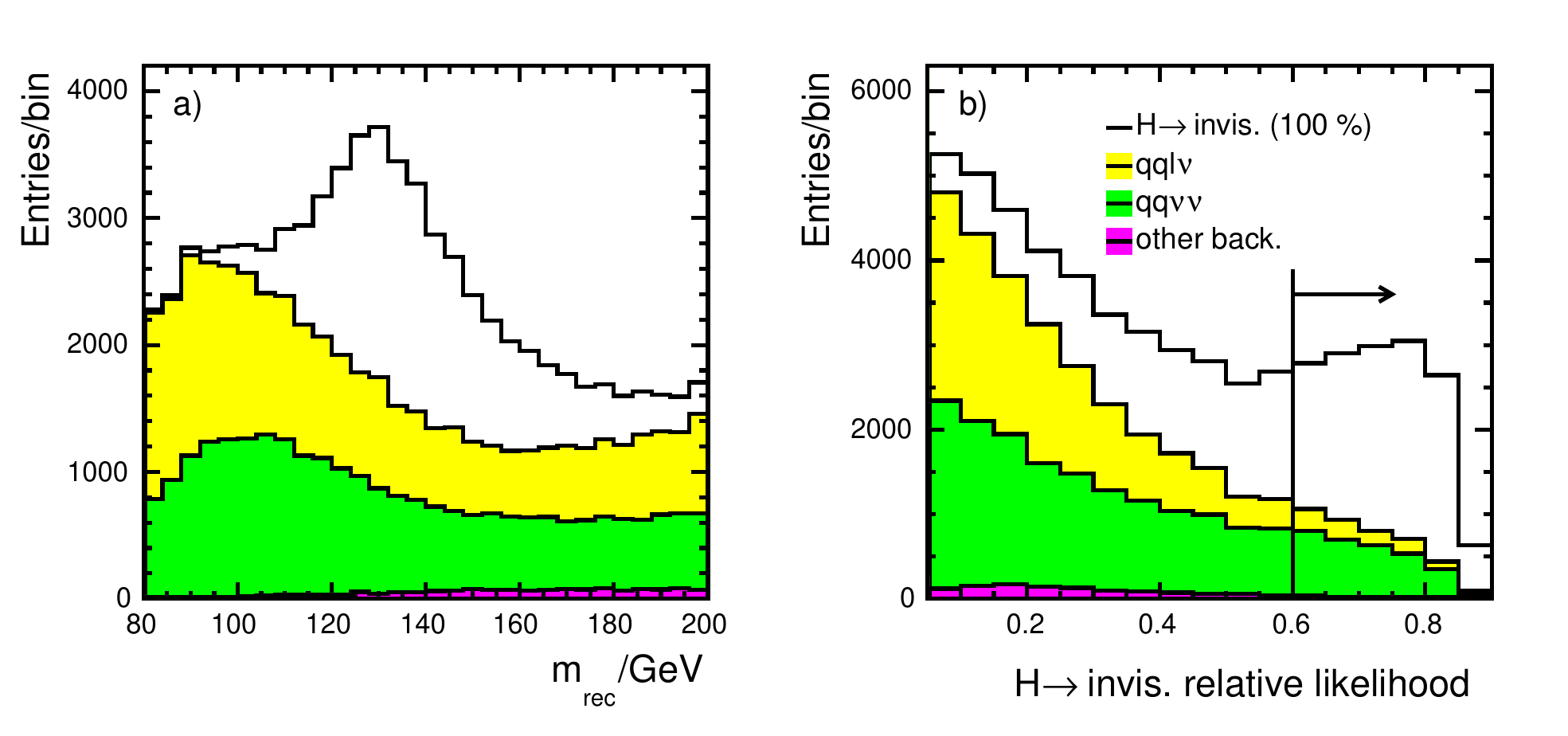} 
      \caption{\label{fig:invisSel} a) The reconstructed hadronic recoil mass distribution for events passing the preselection cuts in the clear two-jet
         topology. b) The invisible Higgs decay relative likelihood distribution for signal and background. In both distributions the event rates are normalised to
         a CLIC integrated luminosity of $500\,\fbinv$ at $\roots=350\,\GeV$. The $\PH\PZ$ signal is shown for the artificial 
         case of $\BR(\PH\to\text{invis.}) = 100\,\%$. The optimal likelihood cut at ${\cal{L}}=0.60$ is indicated by the arrow.}
\end{centering}
\end{figure*}}
In the limit where the $\PH\to\text{invis.}$ branching ratio is small (as expected), 
the expected uncertainty on the number of invisible Higgs decays selected by a particular likelihood cut is driven by the statistical fluctuations on the number of 
background events, $\sqrt{B}$. In this limit, the corresponding uncertainty on the cross section for $\PH\PZ$ production with $\PH\to\text{invis.}$ is given by 
\begin{equation*}
    \Delta \sigma_\text{invis.}    =  \frac{\sqrt{B}}{S} \sigma_{\PH\PZ}^{\rm SM} \,,
\end{equation*}
where $S$ is the number of signal events that would have been selected for the case of a 100\,\% branching 
fraction for $\PH\to\text{invis.}$ This uncertainty is minimised for a relative likelihood cut of ${\cal{L}}>0.60$,
resulting in a $\pm0.58\,\%$ statistical uncertainty on a $\sigma_\text{invis.}$, relative to the SM cross section for
$\epem\to\PH\PZ$. The corresponding selection efficiencies are shown in Tab.~\ref{tab:invisSel}, where 
the expected background from SM $\PH\PZ$ production includes the $\PH\to\PZ\PZ^*\to\PGn\PAGn\PGn\PAGn$ component
that has a SM branching fraction of 0.1\,\%.    

{
\begin{table}[tb]
  \centering
   \begin{tabular}{lrrrr}
    \toprule 
         Process                                & \tabt{$\sigma/\text{fb}$} & \tabt{$\varepsilon_\text{presel}$} & \tabt{$\varepsilon^\text{invis.}_{{\cal{L}}>0.60}$} & \tabt{$N_{{\cal{L}}>0.60}$} \\ \midrule
    $\PQq\PAQq\Pl\PGn$                              &  5914      &    0.9\,\%                       & 0.03\,\%       & 951 \\   
    $\PQq\PAQq\PGn\PAGn$                         &  325       &    14.8\,\%                      &  1.83\,\%       &     2985  \\    
     $\PH\PGne\PAGne$                                 &    52            &     5.6\,\%                 &    0.37\,\%     &   95       \\
    $\PH\PZ$; $\PZ\to\PQq\PAQq$                 &   93      &    0.2\,\%                        &    0.06\,\%     & 31  \\
  \midrule
  $\PH\to\text{invis.}$     \, (100\,\%)                                 &   [93]     &        48.6\,\%                 &   23.52\,\%  & 10983  \\                                                                                                                                                                                   
    \bottomrule
  \end{tabular}
   \caption{Summary of the CLIC $(\PH\rightarrow\text{invis.})(\PZ\rightarrow\qq)$ event selection at $\roots=350\,\GeV$, giving the raw cross sections, preselection efficiency, overall selection efficiency for a likelihood cut of  ${\cal{L}}>0.60$ and the expected numbers of events passing the 
  event selection for an integrated luminosity of $500\,\text{fb}^{-1}$ and unpolarised electron and positron beams. The numbers shown for the invisible Higgs decay modes correspond to a 100\,\% branching ratio. 
 \label{tab:invisSel}}
\end{table}
}

A more optimal approach to extracting the signal cross section is to fit the shape of the likelihood distribution of Fig.~\ref{fig:invisSel}b, 
rather than  simply imposing a single likelihood cut. In the limit that the invisible branching ratio is small, the resulting Gaussian uncertainty on 
the $\PH\PZ$ production cross section with invisible Higgs decays is
\begin{equation*}
   \frac{ \Delta \sigma_\text{invis.}}{\sigma_{\PH\PZ}^{\rm SM}}    =  \pm 0.56\,\% \,,
\end{equation*}
relative to the SM $\epem\to\PH\PZ$ cross section. 
For the SM Higgs, the corresponding expected 90\,\% confidence level upper limit on the invisible Higgs branching ratio is
\begin{equation*}
     \BR(\PH\to\text{invis.}) < 0.94\,\%\ \ \text{at} \  90\,\%  \ \text{C.L.}  
\end{equation*}

\subsection{Model independence of the hadronic recoil mass measurement}

By combining the two analyses for $\PH\PZ$ production where $\PZ\to\qqbar$ and the Higgs decays either to visible or invisible final 
states,
\begin{equation*}
      \sigma(\PH\PZ) = \frac{\sigma_\text{vis.} + \sigma_\text{invis.}}{\BR(\PZ\to\qqbar)} \,,
\end{equation*}
it is possible to determine the absolute $\epem\to\PH\PZ$ cross section in a nearly model-independent manner.
Since the fractional uncertainties (relative to the total cross section) on the visible and invisible cross sections are 1.7\,\% and 0.6\,\% 
respectively, the fractional uncertainty on the total cross section will be the quadrature sum of these two fractional uncertainties, namely
\begin{equation*}
   \Delta\sigma(\PH\PZ) = \pm1.8\,\%\,. 
\end{equation*}
Thus, the \higgsstrahlung cross section can be measured with a precision of better than 2\,\% at $\roots = 350\,\GeV$ using the
hadronic recoil mass (with unpolarised beams).  Such a measurement is competitive with that obtainable from the leptonic recoil mass 
measurement at $\roots =250\,\GeV$, where a precision of $\pm 2.6\,\%$\cite{bib:ILC_Physics_TDR} is achievable with 250\,\fbinv of data 
(assuming $-80\,\%$ and $+30\,\%$ polarisation of the electron and positron beams). The strongest physics argument for 
operating a linear collider at  $\roots =250\,\GeV$ is the model-independent measurement of $\sigma(\PH\PZ)$ that provides a determination of $\gHZZ$.  
If it can be argued that the hadronic recoil mass measurement is effectively independent of the nature of the Higgs boson decay modes 
(including possible extensions to the SM), then the arguments for operating an $\epem$ linear collider at $\roots\sim250\,\GeV$ are greatly reduced; 
almost all other measurements of the properties of the Higgs boson are found to benefit from higher centre-of-mass 
energies~\cite{bib:ILC_Physics_TDR}. In addition, operating at 
$\roots\sim350\,\GeV$ allows the study of Higgs production through the $\PW\PW$-fusion process and the pair production of top quarks.
    
However, the hadronic recoil mass measurement of $\sigma(\PH\PZ)$ can only be truly model 
independent if the overall (visible + invisible) selection efficiency is independent of the Higgs decay mode. 
Tab.~\ref{tab:MIefficiencies} summarises the combined selection efficiency for  $\epem\to\PH\PZ(\PZ\to\qqbar)$, 
broken down into the different Higgs decay modes. Also shown are the 
efficiency for $\PH\to\PW\PW^*$ decays broken down into the different $\PW$ decay modes, 
covering a very wide range of event topologies, from four-jet final states $(\qqbar\qqbar)$ to final states with 
two relatively soft particles, for example the visible tau decay products from $\PW\PW^*\to\PGt\PGn\PGt\PGn$.
For all final-state topologies, the combined (visible + invisible) selection efficiency lies between 19\,\% and 26\,\% compated
to the mean selection efficiency of $\sim23\,\%$; a relative variation of $\pm15\,\%$. It should be noted that these numbers are only indicative, since the measured cross sections are
extracted from fits to the likelihood distributions, rather than from a selection imposing hard cuts. 
\begin{table}[tb]
  \centering
    \begin{tabular}{lrrc}
    \toprule 
         \tabt{Decay mode}    &         \tabt{$\varepsilon^\text{vis.}_{{\cal{L}}>0.65}$} &   
           \tabt{$\varepsilon^\text{invis.}_{{\cal{L}}>0.60}$}    & 
         $\varepsilon^\text{vis.} + \varepsilon^\text{invis.}$        \\  
  \midrule       
    $\PH\to\text{invis.}$  
                                                                     & <0.1\,\%  &     23.5\,\%  & 23.5\,\%      \\
    $\PH\to\qqbar/\Pg\Pg$  
                                                                     &  22.6\,\% &      <0.1\,\% & 22.6\,\%    \\
    $\PH\to\PW\PW^*$  
                                                                     &  22.1\,\% &      0.1\,\%  & 22.2\,\%      \\
    $\PH\to\PZ\PZ^*$  
                                                                     & 20.6\,\% &     1.1\,\%    & 21.7\,\%     \\
    $\PH\to\tptm$  
                                                                     & 25.3\,\% &    0.2\,\%     &  25.5\,\%    \\       
    $\PH\to\PGg\PGg$  
                                                                     & 25.7\,\% &    <0.1\,\%   & 25.7\,\%  \\                                                                            
    $\PH\to\PZ\PGg$  
                                                                     & 18.6\,\% &     0.3\,\%    & 18.9\,\%    \\           
\midrule
        $\PH\to\PW\PW^*\rightarrow\qqbar\qqbar$  
                                                                     & 20.8\,\% &     <0.1\,\%   &   20.8\,\% \\     
        $\PH\to\PW\PW^*\rightarrow\qqbar\Pl\PGn$  
                                                                     & 23.3\,\% &    <0.1\,\%    &  23.3\,\%  \\   
        $\PH\to\PW\PW^*\rightarrow\qqbar\PGt\PGn$  
                                                                     & 23.1\,\% &      <0.1\,\%  & 23.1\,\%   \\        
        $\PH\to\PW\PW^*\rightarrow\Pl\PGn\Pl\PGn$  
                                                                     & 26.5\,\% &   0.1\,\%       &  26.5\,\%  \\   
       $\PH\to\PW\PW^*\rightarrow\Pl\PGn\PGt\PGn$  
                                                                     & 21.1\,\% &   0.5\,\%       &  21.6\,\% \\        
        $\PH\to\PW\PW^*\rightarrow\PGt\PGn\PGt\PGn$  
                                                                     & 16.3\,\% &   2.3\,\%       &  18.7\,\%    \\                                                                                                                                                                                                                                                                                                                                                                                          
    \bottomrule
  \end{tabular}
  \caption{Summary of the efficiencies  of the $\PH(\PZ\rightarrow\qqbar)$ analyses at 
  $\roots=350\,\GeV$, giving the overall selection efficiency for the visible analysis (${\cal{L}}>0.65$) and the invisible Higgs analysis (${\cal{L}}>0.60$). Here $\Pl$ refers to either $\Pe$ or $\PGm$.
 \label{tab:MIefficiencies}}
\end{table}

\begin{table}[tb]
  \centering
   \begin{tabular}{lrc}
    \toprule 
         \tabt{Decay mode}    &         $\Delta$(\BR) &   $\sigma^{\text{vis.}} + \sigma^{\text{invis.}}$ Bias \\  
  \midrule       
    $\PH\to\text{invis.}$       & $+5\,\%$  &    $-0.01\,\%$     \\
    $\PH\to\qqbar$              &  $+5\,\%$ &    $+0.05\,\%$    \\
    $\PH\to\PW\PW^*$        &  $+5\,\%$ &   $-0.18\,\%$      \\
    $\PH\to\PZ\PZ^*$         &  $+5\,\%$ &     $-0.30\,\%$     \\
    $\PH\to\tptm$                & $+5\,\%$   &   $+0.60\,\%$    \\                                                                         
     $\PH\to\PGg\PGg$      & $+5\,\%$   &    $+0.79\,\%$    \\     
    $\PH\to\PZ\PGg$          & $+5\,\%$  &     $-0.74\,\%$    \\                
    $\PH\to\PW\PW^*\rightarrow\PGt\PGn\PGt\PGn$  & $+5\,\%$ & $-0.98\,\%$    \\                                                                                                                                                                                                                                                                                                                                                                                          
    \bottomrule
  \end{tabular}
   \caption{Biases in the extracted $\PH(\PZ\rightarrow\qqbar)$ cross section for cases where the Higgs BR to a specific final state is increased by 5\,\%, i.e. $\BR(\PH\to X) \rightarrow 
  \BR(\PH\to X) + 0.05$.
 \label{tab:MIbiases}}
\end{table}

To assess the impact of the different sensitivities to the different $\PH$ decay topologies, the different Higgs decay modes 
in the $\PH\PZ$ MC samples are reweighted to correspond to modified (non-SM) branching fractions and the total (visible + invisible) 
cross section is extracted as before (assuming the SM Higgs branching ratios). Tab.~\ref{tab:MIbiases} shows the resulting biases in the extracted total 
cross section for the case when a $\BR(\PH\to X) \rightarrow \BR(\PH\to X) + 0.05$. 
In all cases, the resulting biases in the extracted total $\PH\PZ$ cross section are less than 1\,\%, which should be compared to the 1.8\,\% statistical 
uncertainty. These variations represent large deviations from the SM which would be observable in studies of exclusive final states. 
For example, for an integrated luminosity of 500\,\fbinv, a 5\,\% (absolute) increase in branching ratio would   
result in an increase of 3350 $\PH\PZ$ events in that particular Higgs decay topology, including an increase of 230 events with 
either $\PZ\to\mpmm$ and $\PZ\to\epem$ decays. Such large effects would be observable at a linear collider either 
through their impact on exclusive Higgs branching ratio analyses or they would manifest themselves as large excesses of events
in the event samples obtained from the $\PZ\to\epem/\mpmm$ recoil mass analysis. It is therefore reasonable to conclude that
unless very large BSM effects had been previously discovered, the hadronic recoil mass study gives an effectively 
model-independent measurement of the $\PH\PZ(\PZ\rightarrow\qqbar)$ cross section.

\section{The Hadronic Recoil Mass Measurement at the ILC}

\label{sec:ilc}
\begin{figure}[bht]
\centering
      \includegraphics[width=0.425\textwidth]{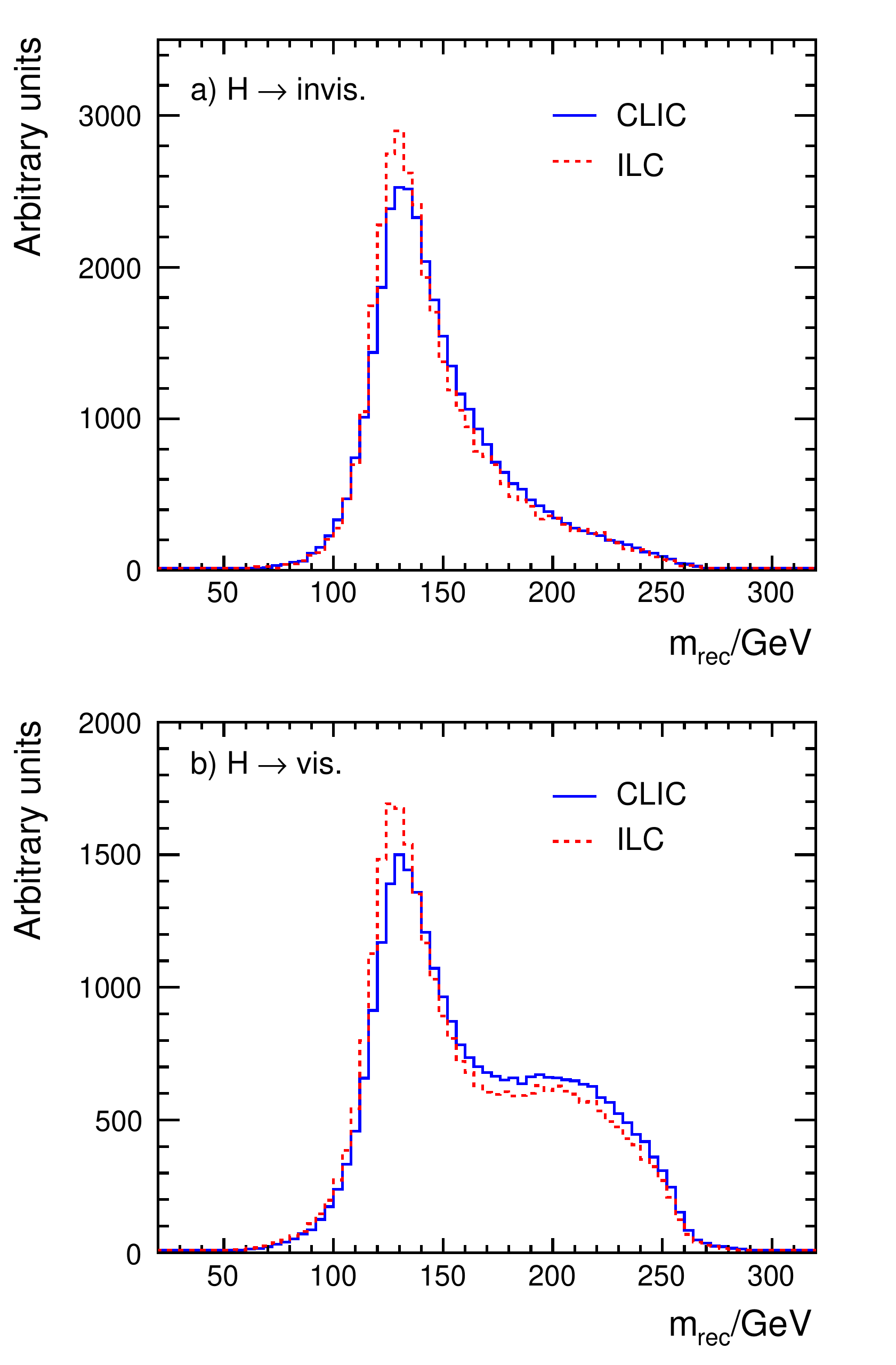} 
      \caption{\label{fig:ILCCLIC} The reconstructed hadronic recoil mass distributions for events passing the preselection cuts  a) for the clear two-jet
         topology of the invisible Higgs decay analysis and b) for the visible Higgs decay analysis. In both cases the distributions compare the CLIC and
         ILC simulations for $500\,\fbinv$ at $\roots=350\,\GeV$, with unpolarised beams.}
\end{figure}

The hadronic recoil mass study presented in this paper was first performed in the context of the CLIC accelerator, where the first stage of the 
machine was assumed to operate at $\roots=350\,\GeV$. The study was then repeated for the ILC at $\roots=350\,\GeV$. Again a full \geant simulation of the detector 
response and a full reconstruction of the simulated events was performed. Since both studies used the same simulation and reconstruction software,
only small differences in precisions on $\sigma(\PH\PZ)$ from the hadronic recoil mass measurement at the ILC and CLIC are 
expected. There are two main effects. Firstly, because of the smaller beam spot at CLIC, the impact of beamstrahlung is greater than for the ILC, 
leading to a larger number of events towards lower values of $\roots^\prime$ at CLIC compared to the ILC, where $\roots^\prime$ is 
the effective centre-of-mass energy of the colliding electron and positron after the radiation of beamstrahlung photons, although the difference
is not large at $\roots=350\,\GeV$.  Secondly, the ILD detector concept used for the ILC studies has more complete calorimeter coverage
down to low polar angles than the CLIC\_ILD detector concept used for the CLIC studies. Both effects will tend to degrade the hadronic recoil mass 
reconstruction for the CLIC configuration compared to the ILC. However, the impact is not large, as can be seen from Fig.~\ref{fig:ILCCLIC}. 

\begin{table}[htb]
  \centering
   \begin{tabular}{lc|ccc}
    \toprule 
     & ${P}(\Pe^-,\Pe^+)$  &  $\Delta\sigma_{\text{vis.}}$ & $\Delta\sigma_{\text{invis.}}$ & $\Delta \,\sigma(\PH\PZ)$ \\  
  \midrule       
    CLIC 500\,\fbinv  \!\!\!\!       &      0, 0         & $\pm1.71\,\%$    & $\pm0.56\,\%$   & $\pm1.80\,\%$   \\ 
    ILC    \,\, 500\,\fbinv \!\!\!\!   &      0, 0         & $\pm1.57\,\%$    & $\pm0.48\,\%$   & $\pm1.63\,\%$   \\  
    ILC   \,\, 350\,\fbinv  \!\!\!\!  & $-0.8,+0.3$ & $\pm1.68\,\%$   & $\pm0.52\,\%$     & $\pm1.76\,\%$   \\                                                                                                                                                                                                                                                                                         
    \bottomrule
  \end{tabular}
   \caption{Summary of the statistical precision achievable on $\sigma(\PH\PZ)$ from the hadronic recoil mass analysis at $\roots=350\,\GeV$ for 
     CLIC and the ILC. The ILC numbers are shown for both zero and the nominal beam polarisations. 
 \label{tab:comparison}}
\end{table}
Tab.~\ref{tab:comparison} compares the statistical precision achievable at a centre-of-mass energy of $\roots=350\,\GeV$ for: 500\,$\fbinv$ at
CLIC with unpolarised beams; 500\,$\fbinv$ at the ILC with unpolarised beams; and 
350\,$\fbinv$ at the ILC with the nominal ILC beam polarisations of $P(\Pe^-,\Pe^+) = (-0.8,+0.3)$. For the same integrated luminosity and unpolarised beams, the precision achievable at the ILC  is approximately 8\,\% better than that at CLIC, reflecting the slightly better recoil mass resolution at the ILC seen in Fig.~\ref{fig:ILCCLIC}. 
Since the instantaneous luminosity at the ILC is expected to scale with the Lorentz boost of the colliding beams $\gamma_{\Pe}$, the time taken to accumulate
350\,$\fbinv$ of data at $\roots$ is comparable to the time required for 250\,$\fbinv$ at $\roots=250\,\GeV$. Hence, for
the nominal ILC beam polarisation of $P(\Pe^-,\Pe^+) = (-0.8,+0.3)$, the statistical precision of 
$1.8\,\%$ achievable on the $\PH\PZ$ cross section at $ \roots=350\,\GeV$ using $\PZ\to\qq$ is directly comparable to the statistical precision 
of 2.6\,\%~\cite{bib:ILC_Physics_TDR, bib:ILD_LOI, bib:Li} achievable with $250\,\fbinv$ of data at $\roots=250\,\GeV$ 
using $\PZ\to\lplm$ decays. This conclusion weakens the motivation for operating a future linear collider significantly 
below the top-pair production threshold.

\section{Centre-of-mass Energy Dependence of the Hadronic Recoil Mass Analysis}

\begin{figure*}[htb]
\centering
      \includegraphics[width=0.9\textwidth]{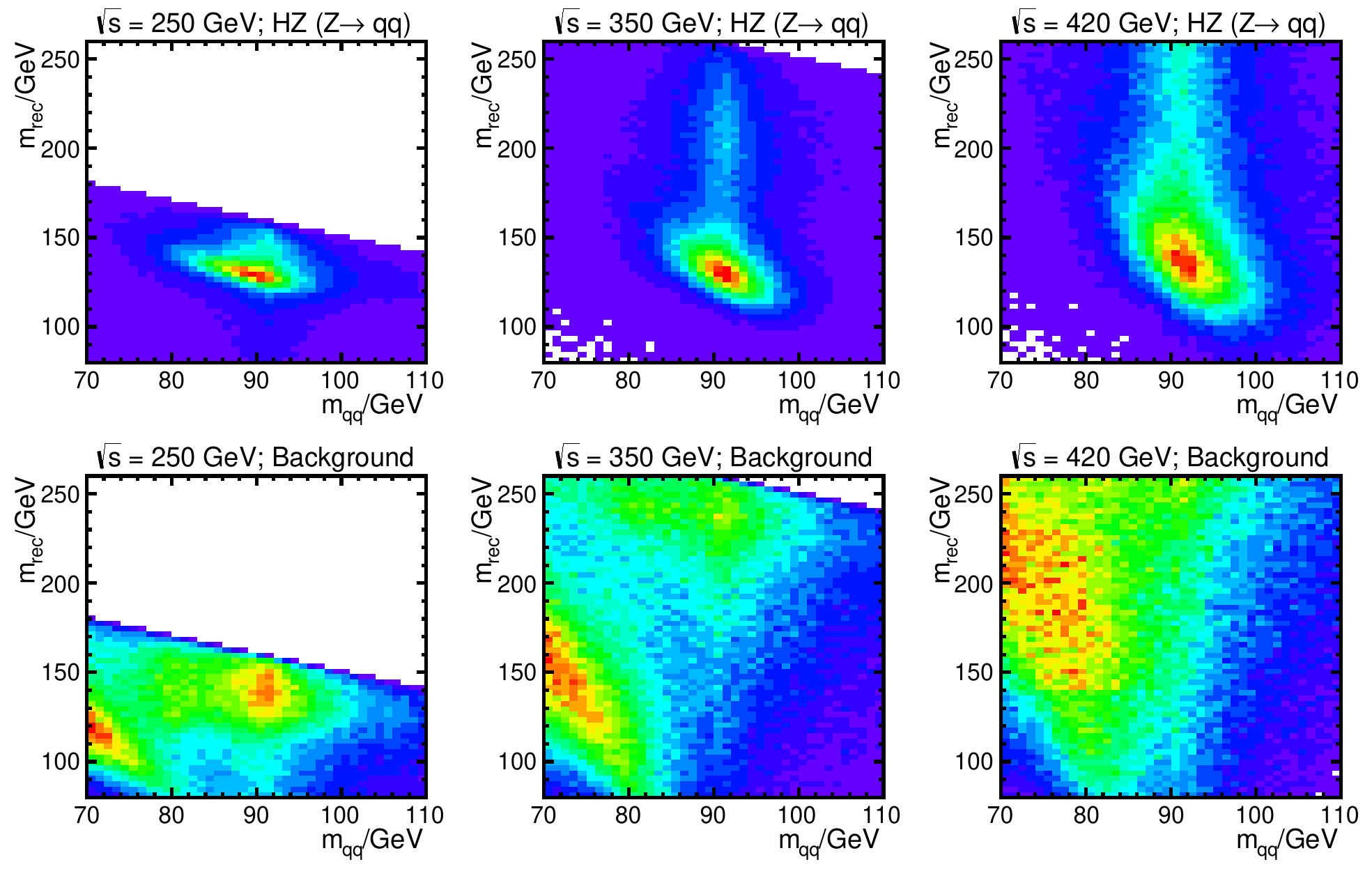} 
      \caption{\label{fig:edep} The two-dimensional distributions of the reconstructed $\PZ$ mass, $\mqq$, versus the reconstructed hadronic recoil mass, $\mrec$, in the visible Higgs decay analysis, broken down into $\PH\PZ(\PZ\to\qq)$ signal and the SM background for CLIC operating at $\roots = 250\,\GeV$,      
      $\roots = 350\,\GeV$ and $\roots = 420\,\GeV$. All events passing the preselection cuts are included. }
\end{figure*}
The hadronic recoil mass analysis described above for $\roots=350\,\GeV$ was repeated for CLIC at $\roots = 250\,\GeV$ and $\roots = 420\,\GeV$. In each case a full set of SM model background  
processes was generated using the \geant simulation of the \clicild detector concept. Because the complete simulation of the CLIC beam is not available for these centre-of-mass energies; the 250\,$\GeV$ samples used that same $\rootsprime/\roots$ distribution as for $\roots = 350\,\GeV$, whereas the 420\,$\GeV$ used the  $\rootsprime/\roots$ 
for the 500\,\GeV option for CLIC. The analysis described in Sect.~\ref{sec:clicanalysis}
was repeated at each centre-of-mass energy using the appropriate distributions for the likelihood function. The binning 
and range used for $\mrec$ in 
the two-dimensional distribution of $\mqq$ versus $\mrec$ was optimized for each centre-of-mass energy. The resulting 
sensitivities are listed in Tab.~\ref{tab:edep}. Compared to $\roots = 350\,\GeV$, the overall sensitivity 
for $\Delta\sigma(\PH\PZ)$ is worse at both $\roots=250\,\GeV$ and $\roots=420\,\GeV$, although for two different reasons (explained below).

\begin{table*}[htb]
  \centering
   \begin{tabular}{lccc|ccc}
    \toprule 
   Machine   & $\roots$ & {$\cal{L}$} & $\sigma(\PH\PZ)$ &  $\Delta\sigma_{\text{vis.}}$ & $\Delta\sigma_{\text{invis.}}$ & $\Delta \,\sigma(\PH\PZ)$ \\  
  \midrule       
    CLIC  & 250\,\GeV & 500\,\fbinv &  136\,fb & $\pm3.63\,\%$    & $\pm0.45\,\%$   & $\pm3.65\,\%$   \\  
     CLIC & 350\,\GeV & 500\,\fbinv &  93\,fb   & $\pm1.71\,\%$    & $\pm0.56\,\%$   & $\pm1.80\,\%$   \\ 
    CLIC  & 420\,\GeV & 500\,\fbinv &  68\,fb  & $\pm2.42\,\%$   & $\pm1.02\,\%$   & $\pm2.63\,\%$   \\                                                                                                                                                                                                                                                                                         
    \bottomrule
  \end{tabular}
   \caption{Summary of the statistical precision achievable on $\sigma(\PH\PZ)$ from the hadronic recoil mass analysis at CLIC for $\roots=250\,\GeV$, 
   $\roots=350\,\GeV$ and $\roots = 420\,\GeV$. In each case unpolarised beams were assumed. 
 \label{tab:edep}}
\end{table*}

Fig.~\ref{fig:edep} shows two-dimensional distributions of $\mqq$ versus $\mrec$, broken down into signal and 
background for the three centre-of-mass energies considered. 
Sect.~\ref{sec:presel}.  For all centre-of-mass energies, the most significant backgrounds are from 
$\epem\to\PQq\PAQq\PQq\PAQq$ and $\epem\to\PQq\PAQq\Pl\PGn$.  The $\PQq\PAQq\Pl\PGn$ background
(predominately from $\epem\to\PW\PW$) accounts for the broad band of events on the left-hand side of the 
background plots. This event population is well separated from the signal region. The more significant 
background arises from the $\PQq\PAQq\PQq\PAQq$ final state, populating the regions with $\mqq \sim \mZ$.  
In this region, the $\PQq\PAQq\PQq\PAQq$ background arises primarily from $\epem\to\PW \PW$, $\epem\to\PZ \PZ^*$ and 
$\epem\rightarrow\PZ\PGg^*$, where the ``$*$'' indicates an off-mass-shell particle; 
the component from $\epem\rightarrow\PZ\PZ$, where both $\PZ$ bosons are on-shell is largely suppressed by the 
preselection cuts. The board recoil mass distribution for the preselected $\PQq\PAQq\PQq\PAQq$ background is 
pushed towards the kinematic limit due to two main effects: i) the pair of jets with the invariant mass closest to $\mZ$ is used to calculate the four-momentum of the assumed $\PZ$ boson, in the case of the $\epem\to\PW\PW\to\PQq\PAQq\PQq\PAQq$ background, this can lead to pairing of two jets from different $\PW$-boson decays; 
ii) for events with significant ISR or beamstrahlung, 
the calculated recoil mass (which uses the assumed centre-of-mass energy $\roots$, rather than $\rootsprime$) is higher than the invariant mass of the recoiling system.

From Fig.~\ref{fig:edep} it can be clearly seen that the width of the recoil mass distribution for 
$\PH\PZ (\PZ\to\qq)$ events increases with increasing centre-of-mass energy. This can be understood from the expression for the recoil mass:
\begin{align*}
    m^2_\text{rec} &= (\roots - E_{\PZ})^2 -(-\vec{p}_{\PZ})^2 \\
                            &= s -2\roots\, E_{\PZ} + E_{\PZ}^2 - \vec{p}_{\PZ}^2 \\
                            &\approx s + \mZ^2 -2\roots\,(E_1+E_2)\,,
\end{align*}
where $E_1$ and $E_2$ are the energies of the two jets forming the reconstructed $\PZ$ boson and assuming $E_{\PZ}^2 - \vec{p}_{\PZ}^2 \approx \mZ^2$, which is true for the signal region. Propagating the errors on the jet energy measurements, $\sigma_1$ and $\sigma_2$, implies that
\begin{align*}
    \sigma_{m_\text{rec}} &= \frac{\roots}{m_\text{rec}}\,\left( \sigma_1^2+\sigma_2^2 \right)^{\frac{1}{2}} \,. 
\end{align*}
Therefore, the recoil mass resolution is expected to worsen with increasing centre-of-mass energy due to 
both the $\roots$ dependence and the fact that the absolute uncertainty on the jet energies 
increases with jet energy ($\sigma_E \sim 0.03E$) and therefore with centre-of-mass energy. The increasing width of the recoil mass 
distribution accounts for the increase of $\Delta\sigma_{\text{invis.}}$ with $\roots$, listed in Tab.~\ref{tab:edep}, 
and the larger value of $\Delta\sigma_{\text{vis.}}$ at $\roots=420\,\GeV$. However, despite the better
recoil mass resolution, the sensitivity to $\Delta\sigma_{\text{vis.}}$ at $\roots=250\,\GeV$ is significantly worse than for 
the other centre-of-mass energies considered. The reason for this can be seen clearly in Fig.~\ref{fig:edep}. At $\roots=250\,\GeV$, $\PH\PZ$ production 
is not very far above threshold and the recoil mass distribution is relatively close to the kinematic limit. 
This is the region populated by the large $\PQq\PAQq\PQq\PAQq$ background passing the preselection cuts, resulting in a greatly reduced separation between 
signal and background in the variable that provides the best distinguishing power, namely $\mrec$.

\section{Summary and Conclusions}
This paper presents the first detailed study of the potential of the hadronic recoil mass analysis at a future linear collider, both for visible Higgs decay modes
and possible BSM invisible decay modes. By combining the analyses for visible and invisible modes, it is shown 
that the measured $\epem\to\PH\PZ(\PZ\to\qq)$ cross section does not depend strongly on the nature of the Higgs boson decay and thus provides a 
model-independent 
determination of $\gHZZ$. The statistical precision achievable at CLIC operating at $\roots=350\,\GeV$ with $500\,\fbinv$ of data with unpolarised beams is
$\Delta\sigma_{\PH\PZ}\approx \pm 1.8\,\%$. A similar precision is obtained for the ILC with $350\,\fbinv$ and the nominal beam polarisation of 
$P(\Pe^+,\Pe^-) = (+30\,\%,-80\,\%)$. In both cases the branching ratio to invisible decay modes can be constrained to 
$\BR(\PH\to\text{invis.}) < 1\,\%$ at 90\,\% confidence level. It is demonstrated that $\roots = 350\,\GeV$ is likely to be close to the optimal energy for 
the hadronic recoil mass analysis; at lower centre-of-mass energies there is less discrimination between signal and background and at higher centre-of-mass energies the measurement is limited by the worsening recoil mass resolution.

It is often stated that operation of a future $\epem$ linear collider close to threshold ($\roots \sim 250\,\GeV$) is necessary to provide an absolute measurement of the coupling between 
the Higgs boson and the $Z$ boson, $g_{\PH\PZ\PZ}$. This is based on the determination of $g_{\PH\PZ\PZ}$ from the recoil mass analysis 
for $\epem\to\PH\PZ(\PZ\to\lplm)$. 
The results presented in this paper show that, for a comparable running time, a statistically more precise measurement can be 
obtained from $\epem\to\PH\PZ(\PZ\to\qq)$ events at $\roots=350\,\GeV$. This conclusion argues against initial operation of a future linear collider at 
significantly below the top-pair production threshold.

\section*{Acknowledgements}
The author would like to thank: colleagues in the CLICdp collaboration, in particular
Christian Grefe, Philipp Roloff and Andre Sailer for their tireless work in generating the CLIC MC
samples used in this study;  colleagues in the ILD detector concept for generating the ILC MC samples used for the results 
reported in Sect.~\ref{sec:ilc}; Aharon Levy and Lucie Linssen for their valuable comments on the first drafts of this paper;  
Aidan Robson, Sophie Redford and Philipp for their comments on the final drafts of this paper; and the UK STFC and CERN
for their financial support.  

\bibliographystyle{elsarticle-num}

\end{document}